\documentclass[hyper]{JHEP3}

\usepackage{epsfig}
\usepackage{latexsym}
\usepackage{amsfonts}
\usepackage{amsmath}
\usepackage{amsthm}
\usepackage{amssymb}
\usepackage{amsbsy}
\usepackage{multirow}

\def\calk         {{\cal K}}
\def\call         {{\cal L}}
\def\calm         {{\cal M}}

\def\calq         {{\cal Q}}

\newsavebox{\uuunit}
\sbox{\uuunit}
    {\setlength{\unitlength}{0.825em}
     \begin{picture}(0.6,0.7)
        \thinlines
        \put(0,0){\line(1,0){0.5}}
        \put(0.15,0){\line(0,1){0.7}}
        \put(0.35,0){\line(0,1){0.8}}
       \multiput(0.3,0.8)(-0.04,-0.02){12}{\rule{0.5pt}{0.5pt}}
     \end {picture}}

\def\be{\begin{equation}}
\def\ee{\end{equation}}
\def\bea{\begin{eqnarray}}
\def\eea{\end{eqnarray}}

\newcommand{\beq}{\begin{eqnarray}}
\newcommand{\eeq}{\end{eqnarray}}

\def\a{\alpha}
\def\b{\beta}
\def\g{\gamma}
\def\G{\Gamma}
\def\d{\delta}
\def\e{\epsilon}

\def\h{\eta}
\def\l{\lambda}
\def\L{\Lambda}

\def\f{\phi}

\def\m{\mu}
\def\n{\nu}
\def\o{\omega}
\def\O{\Omega}
\def\p{\pi}
\def\r{\rho}

\def\s{\sigma}

\def\pa{\partial}
\def\T{\tau}

\def\to{\rightarrow}
\def\nonu{\nonumber \\{}}
\def\half{{1 \over 2}}
\def\sF{{{ F}\!\!\!\!\hskip.8pt\hbox{\raise1pt\hbox{/}}\,}}
\def\som{{{ \omega}\!\!\!\!\hskip.8pt\hbox{\raise1pt\hbox{/}}\,}}
\def\sJ{{{\rm J}\!\!\!\!\hskip.8pt\hbox{\raise1pt\hbox{/}}\,}}

\def\kQ{\calq}
\def\sw{\upsilon}
\def\rQ{{R}}
\def\cJ{{\chi}}
\def\cP{{{\cal P}}}

\newcommand{\cp}{W{}_+}

\title{Chronology protection and the stringy exclusion principle}
\author{Joris Raeymaekers$^1$, Dieter Van den Bleeken$^2$, Bert Vercnocke$^3$

\\
$^1$ Institute of Physics of the ASCR, v.v.i.\\
Na Slovance 2, 182 21 Prague 8, Czech Republic\\

$^2$ NHETC and Department of Physics and Astronomy, Rutgers University, \\
Piscataway, NJ 08855, USA \\

$^3$ CEA Saclay - DSM/IPhT,\,B\^at. 774\\
91191 Gif sur Yvette Cedex, France\\

}

\abstract{We construct a family of supersymmetric solutions to AdS
supergravity in three dimensions, that correspond to type IIB
seven branes wrapped on an internal S$^3\times$T$^4$. These
solutions are generalizations of the three dimensional G\"odel
universe and have closed timelike curves. We propose a enhan\c{c}on-like
mechanism for excising the closed timelike curve region by
including the effects of additional light degrees of freedom.
These take the form of tensionless
 7-brane probes, effectively described through the
backreaction of a smeared domain wall.  The absence of closed
timelike curves in the asymptotic AdS$_3$ geometries obtained in
this way is shown to be equivalent to a unitarity bound in the
dual CFT, known as the stringy exclusion principle.}

\begin{document}
\section{Introduction}
One of the more dramatic consequences of the intrinsic connection
between gravity and the geometry and causal structure of
space-time, is the possible appearance of closed timelike curves
(CTCs). The textbook example of how even a simple, at first sight
completely physical, distribution of energy momentum can lead to
CTCs, is the metric presented by Kurt G\"odel in 1949
\cite{Godel:1949ga}. His solution to Einstein's equations is a
product of a line and  a three-dimensional spacetime with
nontrivial metric \be
\frac{ds^2}{l^2}=-\left(dt+\mu\frac{r^2}{1-r^2}d\phi\right)^2+\mu\frac{dr^2+r^2d\phi^2}{(1-r^2)^2}\label{godelmetric}.
\ee The latter is a solution of three dimensional gravity with negative
cosmological constant $\L = - 1/l^2$ and a source of pressureless
rotating dust. The parameter $\m$ is greater than one and  is
related to the density of the dust as $\m = 1/(1-\r)$. The metric
can be viewed as timelike stretched $AdS_3$, where $\m$ plays the
role of the stretching factor\footnote{G\"odel's original solution
corresponds to the case $\m = 2$.} \cite{Rooman:1998xf}. It is
easily seen that the $\f$-circles are CTCs in the region $
1/\sqrt{\m} <r \leq 1$. For more details on the G\"odel geometry,
see e.g. \cite{Hawking:1973uf}.

Apart from this famous example, many more classical solutions with
CTCs are currently known, including supersymmetric versions
in supergravity theories, both in 3+1 dimensions as well as in
their higher-dimensional parent theories (examples include
\cite{Gauntlett:2002nw},\cite{Israel:2003cx},\cite{Levi:2009az}). Such spacetimes lead to a variety of
pathologies, both within classical general relativity as well as
for interacting quantum fields propagating on them (see
\cite{Friedman:2008dh} for a review and further references). This
led Hawking to propose the Chronology Protection Conjecture,
stating that regions containing CTCs cannot be formed in any
physical process \cite{Hawking:1991nk}. Although he was able to provide a proof in the classical theory, his assumptions are violated once quantum effects are taken into account. As argued in e.g.
\cite{Kay:1996hj} one expects a fully consistent treatment of chronology protection to require a theory where gravity itself is quantized.

Providing such a quantum description of gravity, the AdS/CFT correspondence \cite{Maldacena:1997re} is a promising tool
for studying chronology protection, at least in asymptotically AdS spaces. As illustrated by the above example, the appearance of CTCs
is often a subtle global or IR feature in the bulk, which, because of UV/IR connection in AdS/CFT \cite{Susskind:1998dq}, can  be  expected to appear as a  more obvious UV problem
in the dual CFT.
This idea seems to be borne out in a number of examples, where the appearance of CTCs in the bulk corresponds to an obvious violation of unitarity bounds in the dual  CFT
 \cite{Herdeiro:2000ap,Caldarelli:2001iq,Caldarelli:2004mz,Dabholkar:2005qs}.
Therefore it seems that, if the underlying quantum gravity theory is unitary, the geometries containing CTCs  are unphysical and cannot be formed in any process.

\begin{figure}[t]
\begin{center}
\includegraphics[scale=0.8]{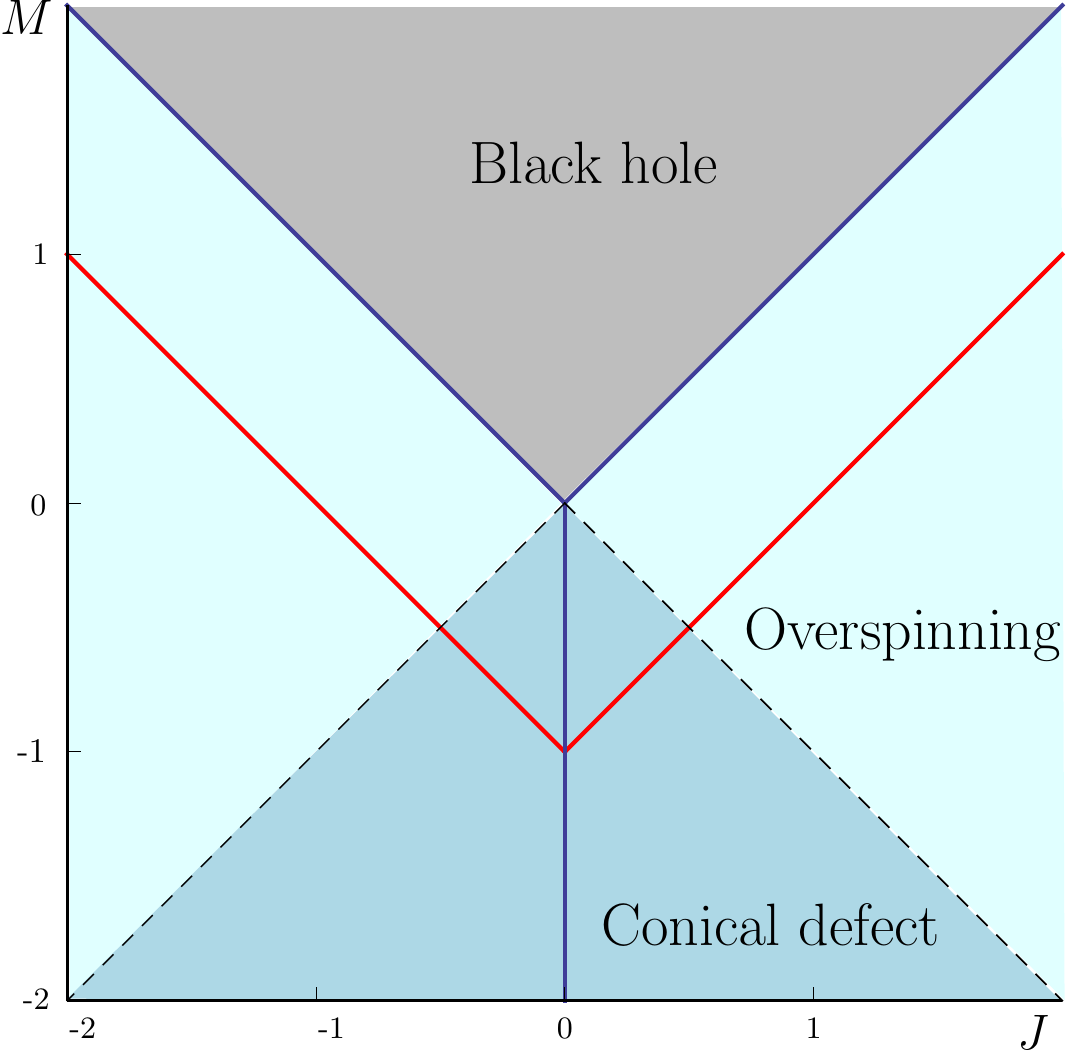}\end{center}\caption{The BTZ metrics (\protect\ref{BTZmetric}) fall in three different classes, depending on the relative values of the mass $M$ and angular momentum $J$.
The metric describes a black hole when $M>|J|$, a naked conical defect when $M<-|J|$ or is free of horizons and singularities when it is overspinning, i.e. $|M|<|J|$.  All metrics above the red lines $M+1=|J|$
have positive conformal weights, while metrics below don't. Asymptotic metrics originating from a matching onto a part of G\"odel space (\protect\ref{godelmetric}) all fall either in the class of the conical defects
or the overspinning class. Considered as complete metrics on their own these spaces have naked CTCs with the exception of ones lying on the blue lines.}\label{MJplane}
\end{figure}

In \cite{Raeymaekers:2009ij}, we demonstrated such a link  between chronology protection and unitarity  in a very simple model where only some very basic features of quantum gravity in asymptotic AdS-space were assumed.
The setup was as follows. We computed the backreaction of a spinning ball of dust, of radius $r_0$ and density $\rho$, in three dimensional gravity. Within the dust ball, the metric is  the G\"odel metric (\ref{godelmetric})
for $r\leq r_0$. Outside of the ball of dust,  where there is no stress-energy, the metric is a generalized BTZ metric  \cite{Banados:1992wn,Banados:1992gq}.
Such metrics depend on two real parameters, the ADM mass $M$ and angular momentum $J$:
\bea
{ds^2 \over l^2} &=&  -  (u -
M) d \tilde t ^2 - J d \tilde t d \tilde \f + u d \tilde \f^2 + {
d u^2 \over 4 f(u)}\,, \label{BTZmetric}\\ f(u) &=&  u^2 -
M u + { J^2 \over 4}\,. \nonumber
\eea
The Israel matching conditions relate the values of $M$ and $J$ to that of the dust ball, $\rho$ and $r_0$. It turns out the for this matched solution $M$ and $J$ always satisfy $|J|>M$, in which case
the metric has no horizons, see figure \ref{MJplane}.
For $u<0$ the $\tilde \phi$ circle in (\ref{BTZmetric}) is
timelike. Closed timelike curves will be present in the combined
spacetime only if the value of the radius $r_0$ where the two
metrics (\ref{godelmetric}) and (\ref{BTZmetric}) are glued
together, exceeds the  critical value $\frac{1}{\sqrt{\mu}}$,
which corresponds to $u_0=0$ in the BTZ metric.

It was our observation in \cite{Raeymaekers:2009ij} that, in the two parameter family of solutions thus  obtained, the solutions with CTCs would precisely correspond to states with negative conformal weight and are hence
excluded in a unitary theory.


In this work, we will  clarify and extend this simple model in
several ways. We start by embedding our model into string theory,
and provide a D-brane interpretation of the G\"odel universe,
building on \cite{Levi:2009az}. We consider type IIB
compactifications on $S^3 \times T^4$, with 7-branes wrapping the
internal manifold. We include in our discussion the 7-branes
producing general SL(2,$\mathbb{Z}$) monodromy that were called
Q7- branes in \cite{Bergshoeff:2006jj}, \cite{Bergshoeff:2007aa}.
The backreaction of these 7-branes produces orbifolds of G\"odel
space which we construct explicitly. These solutions are the
analog of the ones considered in \cite{Bergshoeff:2006jj},
\cite{Bergshoeff:2007aa} in the presence of a negative
cosmological constant, and are D-brane realizations of the
`G\"odel cosmons' discussed in \cite{Banados:2005da}.

In the second part of the paper, we propose a dynamical
mechanism for the
excision of the region in G\"odel space that contains closed timelike curves\footnote{It is well known that G\"odel space is homogeneous and hence closed timelike curves pass through each point. By removing part of the space one can cut these timelike curves before they close. It is in this sense that one can excise a region of G\"odel space so that in the remaining part no timelike curves remain that close on themselves.}. In string theory,
 singularities in the effective supergravity description can  often be shown to be artifacts of the low-energy limit which are resolved
when carefully taking into account the dynamics of the full underlying theory. The mechanism for resolution often involves
the inclusion of nonperturbative degrees of freedom that become light in the problematic region of spacetime and should be included
in the effective description. Such objects will tend to condense at the locus where their tension becomes zero, effectively forming a domain
wall which excises the spacetime singularity. A well-understood example of such a resolution mechanism is the excision of repulson singularities \cite{Johnson:1999qt}.
In that example, the branes that produce the supergravity background are, when considered as probes, found  to become tachyonic in a region of space surrounding
the singularity. On the locus where the tachyonic instability appears, these  branes are tensionless, and thus provide additional massless modes that should be included in the
low energy description.  An effective description of this effect is to consider
 a thin shell of smeared branes on this locus, its backreaction leaves a regular geometry inside the shell and excises the repulson singularity.

When the supergravity solution contains, instead of a curvature
singularity, a region with closed timelike curves, a similar
mechanism was conjectured to apply in \cite{Drukker:2003sc}, and a
five-dimensional example was given\footnote{A somewhat related
mechanism for closed timelike curve removal was proposed in
\cite{Gimon:2004if}. It differs from \cite{Drukker:2003sc} and the example considered here
by the fact that the proposed degrees of freedom responsible for
removing the CTCs were not massless.}. Such examples are very
interesting as they provide hints at what may eventually become a proof of
Hawking's chronology protection conjecture in a UV complete theory
of quantum gravity such as string theory.

We
propose  a dynamical resolution of the closed timelike curve
region of G\"odel space, very much along the lines of \cite{Drukker:2003sc}.
We  begin by identifying Q7-brane probes that become light in a region of G\"odel space. We assume that
they condense at the critical radius where they become exactly massless, forming a
supersymmetric domain wall consisting of branes smeared  on this
radius. We then show that this mechanism is consistent in
supergravity by finding the precise solution on the outside of the
domain wall that satisfies the Israel matching conditions and show that the complete matched spacetime is {\it globally} supersymmetric, i.e. the Killing spinors are analytic through the domain wall. The outside metric
turns out to be a spinning supersymmetric BTZ solution corresponding to chiral primary states in the dual CFT. Our proposal is therefore also
interesting from the point of view of the outside geometry, as it
removes the CTC region of the spinning BTZ metric near its core,
replacing it with a harmless patch of the G\"odel universe.

One new feature of our example is that we  find a discrete  family of such matched solutions, parameterized by the total R-charge they carry.
For a certain range of values of the R-charge, the matched spacetime is free of CTCs.
One very interesting feature, which we take as strong evidence  that
our proposed excision mechanism is consistent, is that this range of R-charges  {\em precisely} reproduces the unitarity bounds on chiral primaries.
That is, not only do we recover the bound that the conformal weight is positive definite, but we also reproduce the upper bound on the conformal weight
which is known as the stringy exclusion principle \cite{Maldacena:1998bw} and is hard to derive from the bulk point of view\footnote{Some indications that such a bound might follow from the supergravity equations were found in \cite{deBoer:2009un}.}.

This paper is structured as follows. First, in section
\ref{secsugra}, we review the (2,0) AdS supergravity theory which
will be our framework, and we discuss the supersymmetric particle
actions that couple to this theory. In section \ref{secsusol} we
then exhibit a number of supersymmetric solutions. There is a
family of solutions corresponding to the backreaction of wrapped
7-branes, all locally G\"odel, and we point out which configuration
corresponds to global G\"odel space. Finally we also review the
supersymmetric vacuum solutions and their relation to chiral
primaries in the bulk. In section \ref{s:DynMech} we present our
main results. First we discuss 7-brane probes and how they can
become massless in the G\"odel background and then we show in
detail how the backreaction of such probes smeared into a
domain wall leads to an asymptotic AdS$_3$ space. Furthermore we show that the combined spacetime is  globally supersymmetric. We then analyze
the discrete family of solutions obtained and show the equivalence
between absence of closed timelike curves and unitarity in the
dual CFT. We conclude and point to some interesting remaining
questions in section \ref{secconc}. For the readers convenience we
included a number of technical appendices. In appendix
\ref{IIBappendix} we detail the reduction of type IIB and its
7-branes to 3 dimensions. In appendix \ref{app:susy} we provide
the details on solving the relevant Killing spinor equations and
in \ref{appmatch} on solving the Israel matching conditions.

\section{AdS supergravity with axion-dilaton}\label{secsugra}
In this section we review the properties of $(2,0)$ AdS
supergravity in three dimensions with an axion-dilaton matter
field. Although we adopt this three dimensional supergravity
point of view throughout most of the paper, our main interest in
this system is as a limit of string theory. The embedding into
string theory requires certain quantization conditions on the
couplings and charges in the supergravity theory. We will keep
track of these quantization conditions along the way, but the
reader interested in purely classical supergravity applications
may safely ignore them.
 We refer to Appendix \ref{IIBappendix} for explicit details on how this theory, and hence its solutions, can be obtained as a consistent reduction and truncation
of Type IIB string theory, note that some more comments on the
U-dualities that connect this setup to Type IIA and M-theory embeddings can be found in \cite{Levi:2009az}.

\subsection{Three dimensional (2,0) AdS supergravity}
It is well known \cite{Achucarro:1987vz, Witten:1988hc} that in
three dimensions Einstein gravity with a negative cosmological
constant can be rewritten as a Chern-Simons theory for the
non-compact gauge group
SL(2,$\mathbb{R}$)$_L\times$SL(2,$\mathbb{R})_R$. One can
generalize this construction to supergravities by replacing the
gauge group by a supergroup. This leads to the ($p,q$) AdS
supergravities of \cite{Achucarro:1987vz, Achucarro:1989gm},
defined as the Chern-Simons theory of the group
OSp($p|2,\mathbb{R})_L\times$OSp($q|2,\mathbb{R})_R$. These are
pure, or minimal, supergravities in that they only contain the
graviton multiplet. The $(4,4)$ and $(4,0)$ cases are familiar, as
they describe the gravity multiplet of the well known AdS/CFT
examples for AdS$_3\times$S$^3$ and  AdS$_3\times$S$^2$
respectively \cite{David:1999nr}. In general one can couple
sigma-model matter to these theories, but to the authors'
knowledge this coupling has not been worked out in detail for
general $(p,q)$. The $(2,0)$ case has however been studied in some
detail \cite{Izquierdo:1994jz}, and as such we find it convenient
to work in this setup as it is the minimal one that contains the
ingredients we need. Note that the embedding in string theory is
most naturally connected to the $(4,0)$ and $(4,4)$ theories, see
appendix \ref{IIBappendix} for some details, but the $(2,0)$
truncation contains the subsector of our interest.

In $(2,0)$ AdS supergravity the gravity multiplet contains not
only the graviton $g_{\m\n}$, but also a $U(1)$ Chern-Simons
gaugefield $A^L$. Furthermore we can couple it to a scalar matter
multiplet, where supersymmetry demands this sigma model matter to
have a target space that is K\"ahler. Denoting the space-time
pullback of the target space vielbein by ${}^*\cP$, the bosonic part
of the action of this theory is
\cite{Izquierdo:1994jz}\footnote{Compared to
\cite{Izquierdo:1994jz} we have changed the metric signature to
(-++), replaced $\g^\a$ by $-i \g^\a$, defined $l = 1/2m$ and
shifted the Chern-Simons gauge field as in their (6.3). Our
$\g$-matrices satisfy $\g_{\a\b} = \e_{\a\b\d}\g^\d$ with
$\e_{012}\equiv 1$.}: \be S_{\mathrm{grav}} = {1 \over 16\pi G}
\int_{\calm_{3}}d^3 x\left[ \sqrt{-g}\left(R+\frac{2}{l^2}
-2(\mu-1){}^*\cP_\a{}^*\bar \cP^{\a} \right)- { 2 \over  l}
\epsilon^{\a\b\g}A^L_\a \pa_\b A^L_\g \right] \,.\label{20action} \ee
Note that we introduced a constant $\mu\geq 1$ to parameterize the
coupling between matter and gravity which, for our embedding in type IIB string theory discussed in appendix
\ref{IIBappendix}, should be taken
 to be\footnote{Note that other values of $\m$ are possible for different embeddings.
For example, $\m=3$ can be obtained in a compactification on a product of two tori with identical complex structures.} $\mu=3/2$.

The fermionic variations under supersymmetry lead to the following
conditions for a bosonic field configuration to be supersymmetric:
\bea
\left( \pa_\a + B^L_{\a\hat \b}\g^{\hat \b}  - {i \over l} A^L_\a - {i\over 2}(\m-1)\;^*\kQ_\a \right)\e &=& 0 \,,\\
(\m-1)\;^*\cP_\a \g^\a \e\ &=& 0\label{20susy}\,. \eea Here we have
introduced the left gravitational Chern-Simons connection $B^L$:
\be B^L_{\a\hat \b} = {1\over 4}\o_{\a}^{\ \hat \g \hat \d}
\e_{\hat \g \hat \d \hat \b} - { 1 \over 2 l}e_{\a
\hat\b}\,.\label{Bdef} \ee Furthermore ${}^*\kQ$ is the pull-back to
space-time of a natural connection defined on the scalar
target space. Given a K\"ahler potential $\calk$, one can
construct the K\"ahler connection $\kQ$ as: \be
\kQ=\frac{i}{4}(\partial\calk-\bar\partial\calk) \,.\ee Finally note
that $\e$ is a complex 2-component spinor and hatted indices are
frame indices.

The case we are interested in is when the scalar target space is
the coset SL(2,$\mathbb{R}$)/SO(2). The vielbein and K\"ahler
connection can in this case be written as \bea
{\cal K} &=& 4 \ln {\rm Im}\tau\\
\cP&=&\frac{1}{\sqrt{2}}\frac{d\tau}{\mathrm{Im}\tau}\\
\kQ&=&\frac{d\mathrm{Re}\tau}{\mathrm{Im}\tau}\label{Qtau}\,, \eea
where $\tau$ is a complex scalar taking values on the upper half
plane: $\mathrm{Im}\tau>0$. The scalar kinetic term in the action becomes:
\be
-2(\mu-1){}^*\cP_\a{}^*\bar \cP^{\a}  = -(\mu-1) \frac{\partial_\a \tau \pa^\a \bar \tau}{({\rm Im} \tau)^2}\,.
\ee
When we think of the 3 dimensional
theory as originating from a compactification of IIB string
theory, see appendix \ref{IIBappendix}, the scalar $\tau$ is the
familiar axion-dilaton of the IIB theory. In string theory the
continues SL(2,$\mathbb{R}$) symmetry is broken by quantization
conditions to the discrete subgroup SL$(2,\mathbb{Z})$.

\subsection{Supersymmetric particles}\label{susypart}
Given the framework of the (2,0)-supergravity, we will be
interested in the supersymmetric charged particles that one can
couple to this theory. A particle in 3 dimensions, just as any
codimension 2 object, will backreact strongly on space-time,
introducing a conical defect proportional to its mass
\cite{Deser:1983tn, Deser:1983dr}. In this subsection we discuss
the source terms corresponding to such point masses, and two ways
to couple them to the other background fields in a supersymmetric
fashion.

\subsubsection{R-particles}
The first type of supersymmetric particle is probably the most
familiar, \cite{Izquierdo:1994jz, David:1999zb,
Balasubramanian:2000rt, Maldacena:2000dr}. It is a particle that
is electrically charged under the U(1) CS gaugefield $A^L$, with a
mass equal to its charge. Since in the dual CFT such particle
states are characterized by non vanishing R-charge, we will refer
to them as R-particles. In the higher dimensional picture the U(1)
gauge symmetry corresponds to a rotation of a sphere factor in the
compactification manifold, and from this point of view the
R-charge is interpreted as angular momentum. The bosonic part of
the action for such supersymmetric particles is simply given by:
\be S_R=-\frac{b}{4G}\int\left( \sqrt{-g}-A^L \right) \,.\ee In an
embedding in string theory the charge $b$ is quantized as $b=R
\frac{4G}{l}=\frac{R}{N^2}$. As detailed in appendix
\ref{IIBappendix}, $N$ is related to the five form flux on the
internal space, and is quantized to be integer, and $R$
corresponds to R-charge in the dual field theory and is also
naturally quantized in integer units.

\subsubsection{Q-particles}\label{Qparticles}
A second type of supersymmetric particle carries magnetic charge
under an axionic scalar of the sigma-model matter.
 Supersymmetry requires that the particle mass depends on the other real scalar comprising the complex axion-dilaton. In terms of the higher dimensional IIB origin of the (2,0) supergravity
(see appendix \ref{IIBappendix}), this type of supersymmetric
particle corresponds to a 7-brane wrapping the complete 7
dimensional internal space. There is a large class of different
supersymmetric 7-branes, classified in the ten dimensional setting
by \cite{Bergshoeff:2006jj}. Each of them is magnetically charged
under a different axionic scalar. The most familiar one is the D7
brane which is magnetically charged under $\mathrm{Re}\tau$, and
produces a monodromy $\tau\rightarrow \tau+1$. As
\cite{Bergshoeff:2006jj} showed, more generally there exists a
supersymmetric 7-brane associated to any SL(2,$\mathbb{Z}$)
monodromy $e^{Q}$, and the different branes, or particles from the
three dimensional standpoint, can thus be classified by the Lie
algebra element: \be Q=\begin{pmatrix}\frac{r}{2}&p\\
-q&-\frac{r}{2}\end{pmatrix} \,.\ee The axion $a_Q$ these particles
couple to is a target space coordinate along the integral curves
of  the Killing vector associated to $Q$: \be
k_{Q}=\mathrm{Re}\left[(p+r
\tau+q\tau^2)\partial_\tau\right]\equiv
\partial_{a_Q}\label{kilvec}\,. \ee Each of these axions $a_Q$ is
most naturally interpreted as the real part of a complex scalar
$\sw_Q$, i.e. $a_Q=\mathrm{Re}\,\sw_Q$. Integrating the condition
(\ref{kilvec}) then gives (assuming $\det Q\geq0$): \be
\begin{cases}
\tau=-\frac{r}{2q}-\frac{\sqrt{\det Q}}{q}\cot\left(\sqrt{\det Q}\,\sw_Q\right)\qquad\mbox{when }q>0\\
\tau=p\sw_Q\ \ \,\quad\qquad\qquad\qquad\qquad\qquad\quad\,\,\,\mbox{when }q=r=0\,.
\end{cases}\label{wQ}
\ee In terms of the new coordinate, a convenient choice of the
target space K\"ahler potential, U(1)-connection and vielbein is
\bea
\calk &=& 4  \ln \sinh \left(2 \sqrt{\det Q} {\rm Im} \sw_Q\right)\,,\\
\calq &=&  {2 \sqrt{\det Q} d{\rm Re} \sw_Q \over \tanh \left(2 \sqrt{\det Q} {\rm Im} \sw_Q\right)}\,,\\
\cP &=&  {  \sqrt{2 \det Q}d\l \over \sinh \left(2 \sqrt{\det Q}{\rm
Im} \sw_Q \right)}\label{lambdapar}\,. \eea

The supersymmetric couplings of a Q-particle were worked out in
\cite{Bergshoeff:2006jj,Bergshoeff:2007aa}. The  tension of the
brane (mass of the particle in three dimensions) is related to the imaginary part of
the new complex scalar $\sw_Q$ : \be T_Q=\frac{p+r
\mathrm{Re}\tau+q |\tau|^2}{\mathrm{Im}\tau}=2\sqrt{\det
Q}\coth\left(2\sqrt{\det Q} \mathrm{Im}\,
\sw_Q\right)\label{tension} \,.\ee The bosonic part of the Q-particle
action is given by: \be S_Q=-\frac{(\mu-1)}{8\p G}\int \left(
T_Q\sqrt{-g}-C_Q \right) \ee where $C_Q$ is the potential of the
fieldstrength $G_Q=dC_Q$, dual to the axion: \be
G_Q=\left(T^2_Q-4\det Q\right)\star\! da_Q \,.\ee It is often useful
to work not in terms of the complex scalar $\tau$, but directly in
terms of the fields the Q-particle most naturally couples to. So
we rewrite the SL(2,$\mathbb{R}$)/SO(2) coset action as an
equivalent action in terms of the scalar tension $T$, and the
gaugefield $C$ (where from now on we will imagine working with a
given charge $Q$ and suppress it as an index): \be
S=-\frac{(\mu-1)}{16\pi G}\int\sqrt{-g}\,dx^3\,\frac{\partial_\mu
T\partial^\mu T+\frac{1}{2}G_{\m\n}G^{\mu\nu}}{T^2-4\det Q} \,.\ee

Finally, let us comment on the relation to the more common
($p,q$)-branes. As the charge matrix $Q$ can be thought of as an
element of the Lie-algebra, it transforms in the adjoint
representation under SL(2,$\mathbb{Z}$) duality. The duality
orbits are the SL(2,$\mathbb{Z}$) conjugacy classes and are
determined by the value of $\det Q$. The requirement that $e^Q$ is an SL(2,$\mathbb{Z}$)-element, leads to the condition that $2 \cos (\sqrt{\det Q}) \in \mathbb{Z}$, or in other words: $\sqrt{\det Q}$ is an integer multiple of either $\pi/3$ or $\pi/2$.
Up to an overall rescaling of $Q$, which labels the quantity of charge, the two physically
different cases that arise in string theory are $\det Q=0$ and
$\det Q=\pi^2$. The case $\det Q <0$ was argued in
\cite{Bergshoeff:2006jj,Bergshoeff:2007aa}   to be unphysical  and
we will not consider it here. The $\det Q=0$ particle corresponds
to a parabolic conjugacy class and contains the D7 brane and all
its SL(2,$\mathbb{Z}$) images, known as ($p,q$)-branes.  The
axion it couples to is generated by an $\mathbb{R}$ subgroup. The
$\det Q=\pi^2$ particle corresponds to an elliptic conjugacy class
and is not not connected to the D7 brane by U-duality. It
couples to an axion  that generates an SO(2) subgroup. This class
contains  objects which can be present in F-theory
compactifications and can be interpreted in terms of bound states
of D7-branes and O7 orientifold planes.
   The most familiar example of this last type, which will also appear in out setup,  is the bound state of 4 D7
branes with an O7 orientifold plane \cite{Sen:1996vd}, which has monodromy $-1$, and corresponds to
$p=q=\pi, r=0$.

\subsubsection{RQ-particles}\label{sss:RQ_part}
Because the R- and Q-particles introduced above are mutually
supersymmetric, one can also write down a supersymmetric action
for a particle that is both charged under an axion, as well as
under the CS gaugefield, which we will refer to as an
`RQ-particle' in what follows. From \cite{Bergshoeff:2005ac,
Izquierdo:1994jz} one can obtain the supersymmetry transformations
of the relevant background bosonic fields, i.e the space-time
dreibein $e^a$, the CS field $A^L$ and $C$, the U(1) dual of the
axion: \bea
\delta_\epsilon e^{a}_\mu&=&i\bar\epsilon\gamma^a\psi_\mu+\mathrm{c.c.}\,,\\
\delta_\epsilon A^L_\mu&=&\bar\epsilon\psi_\mu+\mathrm{c.c.}\,,\\
\delta_\epsilon C_\mu&=&T(\tau)
(\bar\epsilon\psi_\mu+\mathrm{c.c.})+\mbox{terms involving the
scalarino } \l \,.\eea As the three fields all have a variation
proportional to the gravitino $\psi_\mu$, the couplings of the
particle to these three background fields are linked in the
$\kappa$-symmetry analysis of the particle action. The result is
that the relative sign of the R- and Q- charge is fixed for a
supersymmetric particle: \be
S_{RQ}=-\frac{1}{8\p G}\int\Big{(}(\mu-1)T+2\p b)\sqrt{-g}-(\mu-1)C-2 \p bA^L
\Big{)}\label{RQaction} \,.\ee This way of giving R-charge to the
Q-particles might seem a little ad hoc. However, by considering the geometries resulting from the backreaction of such RQ-particles on the metric, we show in
the following section
that for supersymmetric solutions such combined couplings are
actually required. This effect is absent in flat space. It
would be very interesting if there is a more natural worldvolume
explanation of the coupling of the Q-particles to the CS
gaugefield, perhaps through couplings with a worldvolume
gaugefield, or through the curvature of the compactification
manifold \cite{Bergshoeff:2007aa}.

\section{Supersymmetric solutions}\label{secsusol}
In this section we construct supersymmetric solutions to the
(2,0)-supergravity model. We begin by presenting the solutions
that represent the backreaction of general RQ-particle sources. In
particular, we will find that the G\"odel metric discussed in the
Introduction, when accompanied by a suitable Wilson line for the
Chern-Simons gauge field, arises from the backreaction of a
RQ-particle of charge $(\p,\p,0)$. We also show that the backreaction of a `common' D7 brane is a certain orbifold of G\"odel space. We end with a review on which
subset of the BTZ metrics discussed in the Introduction can be
viewed as BPS solutions to  (2,0)-supergravity, and discuss their
interpretation in the dual CFT.

\subsection{Backreacted RQ-particles}\label{ss:Backreacted_RQ}
In \cite{Levi:2009az} it was shown that there exists a family of
supersymmetric solutions to the theory (\ref{20action}). We
shortly review those results here and furthermore consider the
effect of including the supersymmetric sourceterms
(\ref{RQaction}). For a supersymmetric solution it is natural to
make a stationary ansatz for the metric where time is fibered over
a 2-dimensional base manifold: \be
\frac{ds^2}{l^2}=-(dt+\cJ)^2+e^{2\varphi}dzd\bar
z\label{metricansatz}\,.\ee

\subsubsection{Equations of motion} Under this ansatz the equations of
motion simplify drastically. Choosing the source (\ref{RQaction})
to be static with respect to $t$ and located
 at $z=0$, one finds the following equations \cite{Levi:2009az,Bergshoeff:2006jj}:
\bea
\partial\bar\partial\tau+i\frac{\partial\tau\bar\partial\tau}{\mathrm{Im}\tau}&=&-\frac{i\pi}{4}\delta(z,\bar z)(p+r\tau+q\tau^2)\label{taueq}\,,\\
\partial\bar\partial\varphi-e^{2\varphi}&=&-\frac{\mu-1}{4}\frac{\partial\tau\bar\partial\bar\tau}{(\mathrm{Im}\tau)^2}-\frac{1}{4}\left((\mu-1)T+2\p b\right)\delta(z,\bar z)\label{Liouville}\,,\\
d\cJ&=&\frac{i}2 e^{2\varphi}dz\wedge d\bar z\label{eq:J_EOM}\,,\\
\frac{1}{l}dA^L&=&\frac{i\pi}{2}b\delta(z,\bar z)dz\wedge d\bar
z\label{CSequation}\,.\eea These equations have a natural geometric
interpretation in the F-theory embedding of our system. Let us
first discuss  the equations in the absence  delta-function
sources. The first two equations describe the geometry of the
four-dimensional space formed by fibering the F-theory torus over
the base space (where $\tau$ is the complex structure of the torus). The first equation can be solved by a function
$\T$ that is holomorphic, meaning that the torus is elliptically
fibered over the base space. The second equation is a sourced
Liouville equation and  states that the curvature two-form of the
four-manifold is proportional  to the K\"ahler form on the base
manifold. The delta function sources then specify how the
fibration degenerates at the locus of an RQ-particle. The last
equation describes how the time direction is fibered over the
base: it states that the curvature of the  connection $\cJ$ is
precisely the K\"ahler form on the base.

Now we turn to the solution of these equations. We focus here on
solutions describing the backreaction of a single RQ particle and
will not consider multi-particle solutions in this work. We start
with (\ref{taueq}), the equation of motion of the scalar $\tau$.
As was shown in \cite{Bergshoeff:2006jj, Bergshoeff:2007aa} the
most unified description is to work with the scalar $\sw$, defined
in (\ref{wQ}), instead of $\tau$. In terms of this scalar (the
definition of which  depends on $Q$), the solution simply reads:
\be \sw=\frac{i}{2\pi}\log z \ee This form furthermore clarifies
that it is indeed the axion $\mathrm{Re} \sw$ under which the
RQ-particle is magnetically charged. In terms of the standard
axion-dilaton $\T$, this becomes \be
\begin{cases}
\tau=\frac{\tau_0-\bar\tau_0 z^{\sqrt{\det Q}/\pi}}{1-z^{\sqrt{\det Q}/\pi}}\qquad\qquad\mbox{when }q>0\\
\tau=\frac{p}{2\pi i}\log z\,\quad\qquad\mbox{when }q=r=0\,,
\end{cases}\label{soltau}
\ee where $\tau_0=-\frac{r}{2q}+i\frac{\sqrt{\det Q}}{q}$ is the
fixed point of the Killing vector $k_Q$ (\ref{kilvec}).





Next we turn to the equation (\ref{Liouville}) that determines the
conformal factor of the transverse space. The description of RQ particles in AdS
is essentially different compared to
the flat space case. In the flat space case, the $e^{2 \varphi}$
term is absent and the equation becomes a Poisson equation, which
can be solved for all values of the source terms on the right hand
side. In the AdS case however, the only known solution to the
nonlinear  sourced Liouville equation exists in the case that the
$\delta$-function source in equation (\ref{Liouville}) is zero. This
constraint can only be met by turning on on R-charge $b$ for the
source. Evaluating $T$ at $z=0$, i.e. $\tau=\tau_0$, we find that
it implies: \be b=-(\mu-1)\frac{\sqrt{\det Q}}{\pi}
\label{bconstr} \,.\ee
As we mentioned before, it would be interesting
to understand if there is a natural worldvolume argument that
explains this observation. When this constraint is met,  a
solution to this equation was constructed in \cite{Levi:2009az}
for arbitrary holomorphic $\tau$ source and reads \be
e^{2\varphi}dzd\bar z=\frac{\mu}{4}\frac{d\tau
d\bar\tau}{\left(\mathrm{Im}\tau\right)^2}=\mu\frac{\det Qd\sw
d\bar\sw}{\sinh^2\left(2\sqrt{\det Q}\mathrm{Im}\sw\right)} \,.\ee
More explicitly, substituting  the scalar field solution
(\ref{soltau}), the solution for the conformal factor is \be e^{2
\varphi }= {\m \det Q \over \p^2 |z|^{2(1 - \sqrt{\det Q}/\p)} ( 1-
|z|^{2 \sqrt{\det Q}/\p})^2 }\label{solephi} \,.\ee  The geometric interpretation is
as follows: outside of the RQ-brane source, the metric on the base
space is just the pullback of constant negative curvature metric
on target space. At the position of the source ($z = 0$ in our
coordinates), we have a deficit angle $\d$ given by \be \d =  2
(\p  - \sqrt{\det Q}) \,.\ee

Of course, since we turned on a nonvanishing R-charge $b$, this in
turn sources the CS field by equation (\ref{CSequation}), leading
to a Wilson line: \be A^L=-\frac l 2 (\m -1)\frac{\sqrt{\det
Q}}{2\pi}d\,\mathrm{arg} z\label{solA} \,.\ee

Finally let us not forget to present the solution for the
connection $\cJ$ in (\ref{metricansatz}). Note that $\cJ$ is only
determined up to a closed form, for which we will make a
convenient choice. Our solution is \be \cJ = {\mu \sqrt{\det Q}\over \pi} {
|z|^{2  \sqrt{\det Q}/\p} \over  1 - |z|^{2 \sqrt{\det
Q}/\p}} d \arg z\label{solJ} \,.\ee
Note that the solution for the metric and Chern-Simons field when $\det Q = 0$ can be obtained by taking the limit of the above solution.
%
%
%

\subsection{Examples}
Summarized, we found that the backreacted solution of a general RQ
particle satisfying (\ref{bconstr}) at $z = 0$ is given by
 \bea {ds^2\over l^2} &=& -(dt
+ \cJ)^2 + e^{2\varphi} dz d\bar z\,,\nonu
A^L &=& -\frac l2 (\mu -1) \frac{\sqrt{\det Q}}\pi\,d \arg z\,,
\label{eq:GodelSol} \eea
with \be \cJ = {\mu \sqrt{\det Q}\over \pi}\frac{|z|^{2\sqrt{\det
Q}/\pi}}{1 - |z|^{2\sqrt{\det Q}/\pi}}\, d \arg z\,,\qquad
e^{2\varphi} = \frac{\mu \det Q}{\pi^2 |z|^{2(1 - \sqrt{ \det
Q}/\p )}(1-|z|^{2\sqrt{\det Q}/\p})^2}\,. \nonumber \ee
 It was shown in \cite{Levi:2009az} that locally, away from
the source, the metric is that of G\"odel space. As we saw
earlier, the introduction of an RQ-particle  source typically
creates  a deficit angle and can be described as an orbifold of
the global G\"odel metric.  For concreteness, let us now discuss
two specific examples.
\subsubsection{The backreacted D7 particle}
Let us start with the most familiar RQ-particle, namely a pure
D7-brane, which has $(p,q,r) = (1,0,0)$ and zero R-charge
according to (\ref{bconstr}). The metric has deficit angle $2\p$
in the origin and the conformal factor is given by \be e^{2 \varphi} =
{\m \over 4  |z|^2 (\ln |z|)^2}. \ee The base develops  a cusp
singularity or `spike' at the location $z=0$ of the brane $z=0$ : the proper
distance to the brane becomes   infinite but the area of a disc centered on the brane remains finite.

To see that this solution is
an orbifold of G\"odel space, consider the coordinates
\be
x = {\arg z \over 2\p} \qquad y = -{\ln |z| \over 2 \p}\,,
\ee
such that the metric becomes
\be
{ds^2\over l^2 } =  - (dt + \m { dx \over 2 y} )^2 + \m { dx^2 + dy^2 \over 4 y^2}.
\ee
If $x$ would run over the full real line, we would the have global G\"odel metric\footnote{The coordinate transformation to the form \eqref{godelmetric} reads $t \to t + 2 \mu \arg (1 - r e^{i\phi})$ and $x + i y = i \frac{1 + r e^{i\phi}}{1 - r e^{i\phi}}$, see \cite{Levi:2009az} for more details.}, but we see that instead $x$ is to be identified
as
\be
x \sim x + 1.
\ee
Hence the D7-brane solution is the quotient of G\"odel by this action of a $\mathbb{Z}$ subgroup of isometries.

\subsubsection{Global G\"odel space}\label{globalgodel}
Let us now focus on the special case $\sqrt{\det Q}=\pi$, and fix an SL(2,$\mathbb{Z}$) duality frame by choosing $p=q=\pi, r=0$. In this case there is no
deficit angle, and the solution (\ref{eq:GodelSol}) becomes global G\"odel space. It is convenient to choose coordinates $z=r e^{i\phi}$:
\bea
\frac{ds^2}{l^2}&=&-\left(dt+\mu\frac{r^2}{1-r^2}d\phi\right)^2+\m\frac{dr^2+r^2 d\phi^2}{(1-r^2)^2}\,,\nonu
A^L&=&-\frac{l}2 (\mu-1) d\phi\,,\nonu
\tau&=&i\frac{1+re^{i\phi}}{1-r e^{i\phi}}\,. \label{eq:GodelBackground}\eea Often it will also be
useful to express the solution not in terms of $\tau$, but rather
in terms of the tension scalar $T$ and the fieldstrength $G$,
introduced above: \be T=2\p \frac{1+r^2}{1-r^2}\qquad
G=\frac{8\p lr}{(1-r^2)^2}dt\wedge dr+\frac{8\mu l
r^3}{(1-r^2)^3}d\phi\wedge dr\,. \ee As we show in some detail in
appendix \ref{app:susy} this solution is supersymmetric, with
Killing spinors: \be \epsilon=e^{-it}\epsilon_0\,. \ee The
presence of the Wilson line and the dilatino impose the same
projection condition: \be \gamma_{\bar z}\epsilon_0=0\,. \ee Hence
we can think of this global G\"odel solution as the backreaction
of a $p=q=\p,\, r=0,\ b=-(\mu-1)$ particle sitting at the origin
$r=0$. One can compute its charges by integrating the axion
$\mathrm{Im} w=d\phi$, defined in (\ref{wQ}), and the Chern-Simons
gauge field, along a curve enclosing the origin. This brane make
its presence felt through the fact that the fields of type IIB
pick up an SL(2,$\mathbb{Z}$) monodromy $- \bf 1$  when encircling
it,\footnote{Even though the axion-dilaton, which transforms under
PSL(2,$\mathbb{Z}$), is insensitive to this monodromy, other type
IIB fields such as the two-forms do transform under this element
of  SL(2,$\mathbb{Z}$).} as well as trough the Aharonov-Bohm
vortex for the Chern-Simons gauge field. This D-brane
interpretation of the global G\"odel metric will be very useful
when discussing our stringy mechanism for chronology protection in
Section \ref{s:DynMech}.

\subsubsection{Other $\det Q >0$ solutions}
More general $\det Q >0$ particles produce orbifolds of the above
G\"odel metric with fixed point  in the origin, where the brane is
located. For example, the case $p=q=\pi/2, r=0$, which produces an
S-duality monodromy, corresponds to the $\mathbb{Z}_2$ orbifold of
G\"odel. Taking $p= q= r= {4 \p \over 3 \sqrt{3}}$ leads to a
$\mathbb{Z}_3$ orbifold. Such orbifolds were dubbed `G\"odel
cosmons' in the literature \cite{Banados:2005da}, and our
solutions give a D-brane interpretation for  such geometries.

\subsection{Supersymmetric BTZ solutions}\label{susyBTZ}
We are also interested in supersymmetric solutions to the equations of motion of (2,0) supergravity when the axidilaton is constant, $\tau = cst$.  Because of the absence of any stress energy these geometries
are locally AdS$_3$. These solutions are of the generalized BTZ form introduced in the introduction, eq.\ \eqref{BTZmetric}.
We are interested in the metrics in the conical defect or overspinning regimes of parameter space, where $|J|-M\geq 0$.
A discussion of the supersymmetry of the BTZ metrics  in the
context of (2,0) supergravity was already done in
\cite{Izquierdo:1994jz}. We will find it useful to repeat their
analysis in a different coordinate system, adapted to the
case\footnote{By choosing the orientation of $\tilde \phi$ we can
always assume $J>0$, as $\tilde\phi\rightarrow -\phi$ together
with $J\rightarrow -J$ leaves the metric invariant. By this
symmetry all our results derived in the case of $J<0$ have an
analogue for $J<0$, found by changing the sign of $\tilde\phi$.}
$J-M>0$ that will be of our interest. With respect to the
coordinates used in (\ref{BTZmetric}) we define: \bea
t &=& - \a\, \tilde t \,,\\
\f &=& \tilde \f +  \tilde t\,,\label{fitransf}\\
\a U &=& u - {J \over 2}\label{newBTZcoords}\,.
\eea
Where the parameter $\alpha=\sqrt{J-M}$ will play an interesting an natural role later on.
Furthermore, let us allow for the presence of a Chern-Simons Wilson line along the angular direction $\phi$ as in the G\"odel solution. We thus have for the metric and Chern-Simons field:
\bea
\frac{  ds^2}{l^2 } &=&  -  (d  t + U d \f)^2 + \tilde f(U)  d  \f^2 + { d U^2 \over 4 \tilde f(U)}\,,\quad \text {with} \qquad \tilde f(U) = U^2 + \alpha U + J/2
\,,\nonu
A^L &=& \frac l2 \cp d \phi\,,\label{eq:BTZmetric_newcoords}
\eea
where $\cp$ is a constant.
Note that we have chosen the Wilson line for $A^L$ to be purely leftmoving in terms of the asymptotic AdS coordinates (see(\ref{fitransf})), so that it obeys the
standard boundary condition for Chern-Simons fields \cite{Kraus:2006nb}. Note also that the metric (\ref{genBTZ}) remains regular in the extremal limit $\a \to 0$.
In this case the metric represents a near-horizon scaling limit of the extremal BTZ black hole. We will discuss this limit further in section \ref{extremelimit}.

\paragraph{Killing spinors.}
In appendix \ref{app:susy}, we show this solution is supersymmetric with Killing spinors
\be
\epsilon = e^{-it - i\frac n2\phi} \e_0\,, \qquad\text{with}\quad  \g_{\hat{\bar z}} \e_0 = 0 \qquad (\Leftrightarrow \g^{\hat t} \e_0 =  i \e_0) \,.\label{eq:KS_BTZ_2}
\ee
The parameter $n$ is an integer and it determines the periodicity of the spinors. It relates the geometry to the Wilson line parameter as
\be
J - M = (1-\cp - n)^2\,, \qquad n \in \mathbb{Z}\label{eq:JM_W}\,.
\ee
When $n$ is even, the spinors are periodic (corresponding to Neveu-Schwarz or NS boundary conditions) and when $n$ is uneven, the spinors are anti-periodic (Ramond or R boundary conditions).

These spaces preserve half the supersymmetry in (2,0) supergravity, except in the case $\cp = 0$ which preserves all supersymmetry and represents the supersymmetric CFT ground state on the leftmoving side\footnote{Note that since there is only supersymmetry on the left, tensoring the left groundstate with an arbitrary excited state on the right does not break any supersymmetry}. Note that global AdS$_3$ is the special case $M=-1, J = 0$.

The parameters $M,J$ in the BTZ metric are only sensitive to the combination $\cp + n$, implying that the same metric with a different Wilson line corresponds to a different state in a different sector. This map between different states obtained by such an integer shift in the Wilson line goes under the name of spectral flow. As this map is an isomorfism, we can without loss of generality focus on one sector. So throughout much of the paper we will choose $n=0$, putting us in the NS sector, where one has $J-M = (1- \cp)^2$. But it is important to keep in mind that all our statements will have analogies in the other sectors as well, through spectral flow. When appropriate we will again point this out in some more detail.

\paragraph{Dual CFT and unitarity.}
We briefly discuss part of the  AdS/CFT dictionary.  In the dual CFT, BTZ type solutions (\ref{BTZmetric}) with leftmoving Wilson lines represent states with conformal weights and U(1) R-charge given by \cite{Balasubramanian:1999re},\cite{Kraus:2006nb}
\bea
L_0 &=& {c \over 24} ( M - J + 1 + \cp^2 ) \,,\\
R &=& {c \over 12} \cp\,,\\
\bar L_0 &=&\frac{c}{24} (M + J + 1 )
\eea
We find it convenient to use an overall normalization proportional to the central charge $c$ of the CFT, which for any asymptotically AdS$_3$ space is given by \cite{Brown:1986nw}:
\be
c=\frac{3l}{2G}\,.
\ee
The requirement of unitarity in the CFT and the presence of extended supersymmetry, lead to several bounds on the quantum numbers given above. First, the conformal weights should be positive $L_0\geq 0, \bar L_0 \geq0$. Second, supersymmetric states also obey a unitarity bound on the R-charge \cite{Lerche:1989uy}, which in the NS sector takes the form $R \leq c/6$.

We only consider supersymmetric states in the left-moving sector and concentrate therefore on $L_0$ and $R$. In the case that $n=0$, the quantum numbers those quantum numbers become:
\bea
L_0 &=& {c \over 12} \cp\,,\qquad
R = {c \over 12} \cp\,.
\eea
These states that satisfy $L_0=R$ are called chiral primaries, and it is well known that they are the supersymmetric states in the NS sector, so this is nicely consistent with our result that they obeying the supersymmetry condition \eqref{eq:JM_W} with $n=0$.

\begin{figure}
\centering
\includegraphics[width=.5\textwidth]{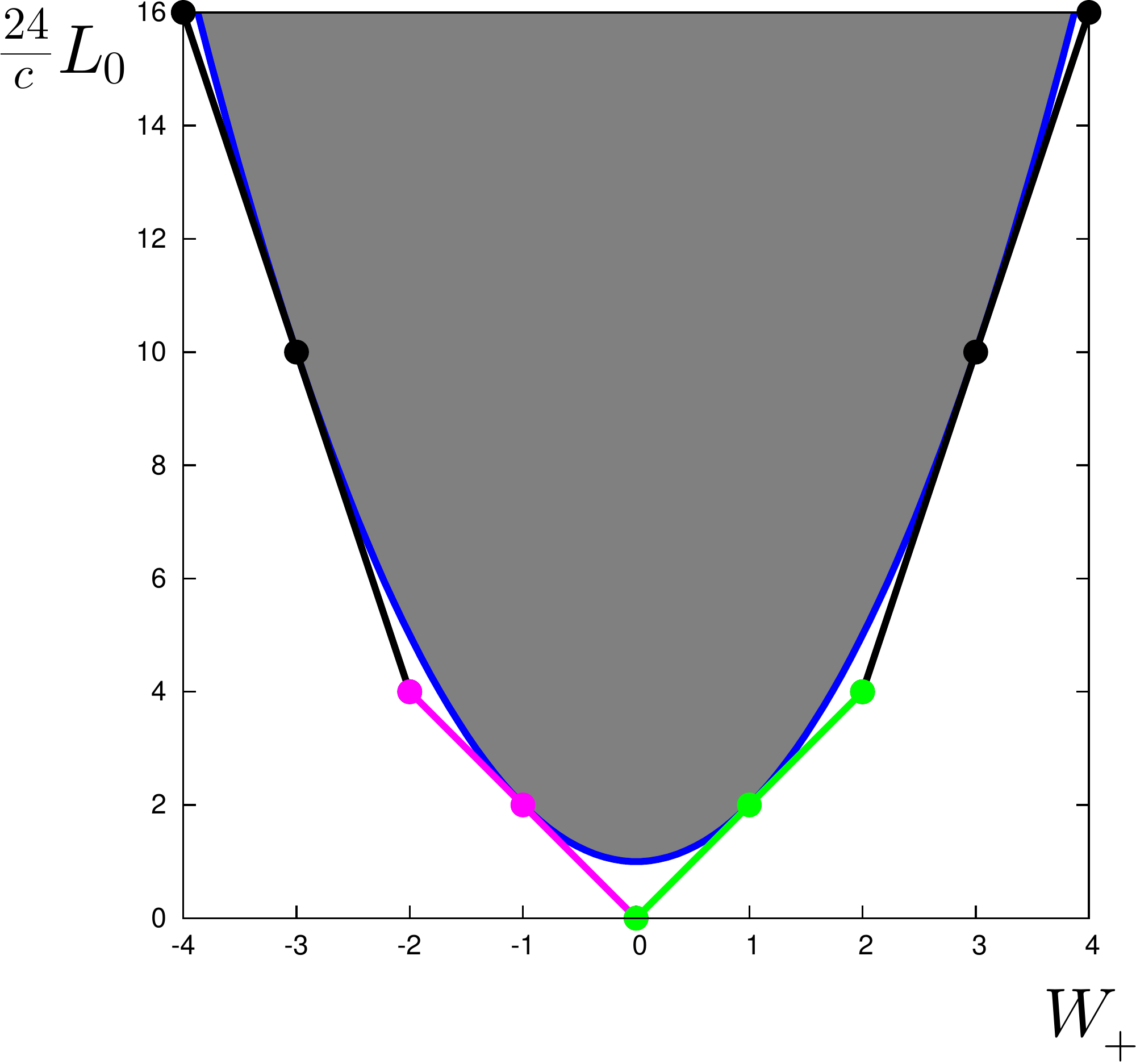}
 \caption{The plane of charges $(L_0, W_+)$ for the BTZ solutions with Wilson line (\protect\ref{eq:BTZmetric_newcoords}). The green line corresponds to the chiral primaries. The lines in other colors are states from other sectors in  the CFT, which can be mapped to the chiral primaries  through spectral flow (i.e. states corresponding to geometries with $n\neq 0$ in (\protect\ref{eq:JM_W}). The parabola corresponds to the black hole threshold $M = J$, states under the parabola have $J>M$ (remember we chose $J>0$). Note that states of middle R-charge in their sector touch on the black hole threshold.}\label{fig:NS_ChiralPrimaries_L0R}
\end{figure}

Note that the unitarity bounds $L_0\geq 0,\, R \leq c/6$ applied to the NS chiral primaries give a condition on the Wilson line
\be
0\leq L_0 = R \leq \frac c 6\qquad \Longrightarrow \qquad  0 \leq\cp \leq 2\,.\label{unitboundCP}
\ee
The allowed unitary CFT states and the corresponding geometries in the $(J,M)$ plane are summarized in figure \ref{fig:NS_ChiralPrimaries_L0R}.

%
%

The upper bound on the R-charge is a result from superconformal field theory \cite{Lerche:1989uy}. States violating this bound are forbidden by unitarity and, according to the AdS/CFT conjecture, cannot be part of the spectrum in a consistent quantum gravity theory on AdS$_3$. It is however not clear at all that such a bound can already be observed in the supergravity approximation. The AdS/CFT conjecture however suggests that in the full string theory this bound will be explictely imposed in the bulk, and for that reason the bound goes under the name of the {\it stringy exclusion principle} \cite{Maldacena:1998bw}. Further on we observe exactly this effect, by taking into account additional light brany degrees of freedom we see how the naive supergravity description gets corrected and how the stringy exclusion principle is explicitly realized directly in the bulk, without resorting to the AdS/CFT correspondence.

Finally, remember that by spectral flow a similar unitarity bound applies in the other sectors in the CFT as well. Table \ref{tab:StringyExclus} provides an overview.

\begin{table}[ht!]
\centering
\begin{tabular}{|c|c|c|ccc|}
\hline
{\bf Sector}&{\bf Boundary conds.}&{\bf Quantum \#s}&\multicolumn{3}{c|}{\bf Unitarity bounds}\\[2mm]
\hline \hline
$n=0$&NS&$L_0 = {c \over 12} \cp$&\multirow{2}{*}{\vspace{-4mm}$0 \leq \rQ \leq {c \over 6}$}&\multirow{2}{*}{\vspace{-4mm}$\Leftrightarrow$}&\multirow{2}{*}{\vspace{-4mm}$ 0 \leq \cp \leq 2$} \\[2mm]
&& $\rQ = {c \over 12} \cp$&&&\\[2mm]
\hline
$n=1$&R&$L_0 = {c \over 24}$&\multirow{2}{*}{\vspace{-4mm}$| \rQ| \leq {c \over 12} $}&\multirow{2}{*}{\vspace{-4mm}$\Leftrightarrow $}&\multirow{2}{*}{\vspace{-4mm}$|\cp| \leq 1 $}\\[2mm]
&&$\rQ = {c \over 12} \cp$&&&\\[2mm]
\hline
general $n$&$n$ even: NS&$L_0 = \frac c {24}[-n(n-2)$&\multicolumn{3}{c|}{\multirow{2}{*}{\vspace{-4mm}$-n  \leq \cp \leq 2 - n $}}\\
&&$\quad\quad-2(n-1)\cp ]$&&&\\
&$n$ uneven: R&$\rQ = \frac{c}{12}\cp$&&&\\
\hline
\end{tabular}
\caption{Quantum numbers and the unitarity bounds following from
supersymmetry in the left-moving sector of the CFT. For $n=0$, we
have chiral primaries in the NS sector, for $n=1$ ground states in
the R sector and for $n=2$, anti-chiral primaries in the NS
sector.} \label{tab:StringyExclus}
\end{table}

\section{A dynamical mechanism for excising closed timelike curves}\label{s:DynMech}

In this section, we will propose a dynamical mechanism for the
excision of the region in G\"odel space that contains closed timelike curves.
We  begin by identifying BPS brane excitations in string
theory that become light in a region of G\"odel space. We assume that
they condense at the critical radius where they become exactly massless, forming a
supersymmetric domain wall consisting of branes smeared  on this
radius. We then show that this mechanism is consistent in
supergravity by finding the precise solution on the outside of the
domain wall that satisfies the Israel matching conditions. This
will turn out to be a spinning supersymmetric BTZ solution of the
type discussed in section \ref{susyBTZ}. Our proposal is therefore also
interesting from the point of view of the outside geometry, as it
removes the CTC region of the spinning BTZ metric near its core,
replacing it with a harmless patch of the G\"odel universe.

One new feature of our example is that we will find a discrete  family of branes that become light in G\"odel, distinguished by
their R-charge quantum number (i.e. their angular momentum on the internal sphere).
By dialling this quantum number we obtain, after working out the construction outlined above,  a family of spinning supersymmetric BTZ geometries
on the outside that are glued to G\"odel in a manner that removes all CTCs. One very interesting feature, which we take as strong evidence  that
our proposed excision mechanism is consistent, is that the family of BPS geometries obtained in this way reproduces {\em precisely} the unitarity bound on BPS states
in the dual CFT, corresponding to the stringy exclusion principle. For example, in the case of the geometries corresponding to chiral primary states in the CFT,
 not only does our construction know about the lower bound on the conformal weight
(which could have been anticipated from the results of \cite{Raeymaekers:2009ij}), but it also correctly reproduces the upper bound. As far as we know, this is the first time
the stringy exclusion principle in AdS$_3$/CFT$_2$ is reproduced from physics in the bulk\footnote{This is in contrast with higher dimensional cases such as the AdS$_5$/CFT$_4$,
where the stringy exclusion bound follows from the physics of giant gravitons \cite{McGreevy:2000cw}. It would be interesting to explore in how far our construction can be seen as the analog of the giant graviton in
AdS$_3$. It might be interesting to note that there are some hints the stringy exclusion principle is reproduced in the physics of multicenter solutions \cite{deBoer:2009un}.}

\subsection{Probing  G\"odel space with RQ particles }
The first step in our proposal involves the identification of stringy degrees of freedom
that become light in a region of G\"odel space. In section \ref{ss:Backreacted_RQ}, we showed that the G\"odel solution arises as the backreacted solution corresponding to an RQ brane in the origin
carrying $(\p,\p,0)$ Q-particle charge as well as R-charge. We look for probe branes that cancel the Q7-charge in the background so as to produce
a space that is asymptotically AdS. To do this in a supersymmetric manner, we need to consider RQ particles carrying $(-\p,-\p,0)$. These branes
have negative tension and their worldvolume action is minus the action for a $(\p,\p,0)$ particle. We furthermore allow them to
carry an amount of R-charge which we call $b$. Let us now demonstrate that such RQ particles  have the desired properties.
The worldline action for such particles was discussed in section \ref{sss:RQ_part}, where it was also shown to be supersymmetric.
It is given by:
\be
S = {1 \over 8 \p G} \left[(\m - 1) \left( \int T\sqrt{-g} - \int C  \right) - 2 \p b \left( \int  \sqrt{-g} - \int  A^L\right) \right]\,.
\ee
Let us now evaluate the worldline action in the G\"odel background (\ref{eq:GodelBackground}), choosing a static gauge with respect to the G\"odel time $t$.
One easily sees that such a particle experiences a flat potential,
which is a consequence of the fact that these probe particles are mutually BPS with the brane that produces the background
geometry.
The  effective mass of the particle is $r$-dependent and  has a positive contribution from
the R-charge part and a negative contribution from the Q7-part:
\be
M_{eff} \sim  b - {(\m -1) \over 2 \p}  T = b - (\m -1 ) { 1 + r^2 \over 1 - r^2}\,.
\ee
Hence we see that, for $b \geq\mu - 1$ which we will assume from now on, these  branes become tachyonic in the region
\be
r^2 > { b - (\m -1) \over b+ (\m -1)}\,.
\ee
At the critical radius
\be
r_0^2 \equiv { b - (\m -1)  \over b+ (\m -1)}\,,\label{wallpos}
\ee
the probe particles become massless, and we expect such probes to form a condensate. We now explore the effects of such a condensate.

\subsection{Consistency of excision}
\begin{figure}
\begin{center}
\includegraphics[scale=0.6]{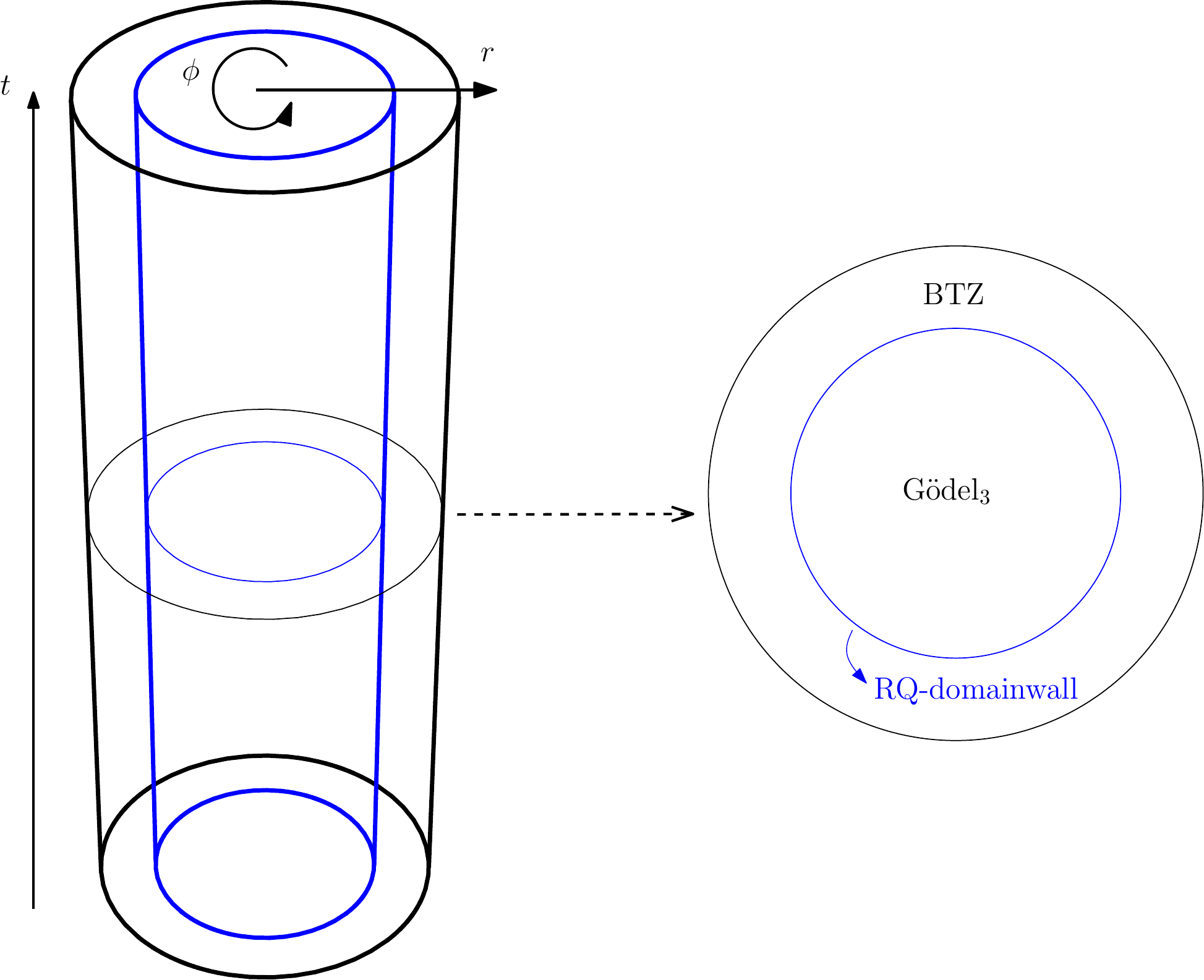}
\caption{Here the global topology of the glued solution is depicted. Inside, i.e. for $0\leq r\leq r_0$, we have the G\"odel solution. A domain wall of smeared RQ-particles interpolates between the outside vacuum geometry, which takes the form of a BTZ metric with specific mass and angular momentum.}
\label{gluingfig}\end{center}
\end{figure}

In a first approximation, we can view such a condensate as a thin shell, formed by the RQ-branes discussed above, smeared
over the critical surface $r = r_0 =  { b - (\m -1)  \over b+ (\m -1)}$ with some constant density $\r$. The shell forms a domain wall separating a
solid cylinder cut out of G\"odel space on the inside,
and a vacuum solution without Q-particle charge outside, see figure \ref{gluingfig}. The values of $\r, b$ and the outside geometry are fixed by the requirement that the jump conditions
on the various fields (i.e. the the Israel matching conditions)
are satisfied for our particular sources on the wall.
This construction will remove the CTCs if
$r_0 \leq 1/\sqrt{\m}$.

\subsubsection{The matched configuration}
We show below that the matched configuration is completely specified by the
total R-charge of the solution, or equivalently by the value of the Chern-Simons Wilson line at infinity
which we will call $W_+$.
Our  matched solution is as follows.

On the inside, have the G\"odel solution produced by a $(\p,\p,0)$ RQ particle at $r=0$ as discussed in section \ref{globalgodel}:
\bea
\frac{ds^2_-}{l^2 } &=&  -(d t+\mu  \frac{ r^2}{1-
r^2}d\phi)^2 + \mu \frac{d r^2+
  r^2d\phi^2}{(1- r^2)^2}\,,\\
T_- &=& 2 \p { 1 + r^2 \over 1 - r^2}\,,\\
G_-  &=&  {8\p  l r \over (1 - r^2)^2 } dr  \wedge \left( dt + {\m  r^2 \over (1 -r^2)} d\f\right)\,,\\
A^L_- &=& W_- {l \over 2} d\f \qquad W_- = 1 - \m\label{inside}\,.\eea
The radial variable runs from 0 to $r_0$, where we place a thin shell of $(-\p,-\p,0)$ particles with R-charge $b$, smeared
with a constant density $\r$. The values of these parameters are
\bea
r_0^2 &=& {W_+ \over 2(\m -1) + W_+} \label{r0sol}\,,\\
b &=&   W_+ + \m -1\label{bsol}\,,\\
\rho &=& {1 \over 2 \p }\,. \label{rhosol}
\eea

On the outside, we take a BTZ-type metric (\ref{eq:BTZmetric_newcoords}) for values of the radial coordinate $U \geq U_0 $, with a  Wilson line for $A^L$ turned on:
\bea  \frac{ds_+^2}{l^2} &=&   -  (d \tilde t + U d\tilde \f)^2 + \tilde f(U)  d \tilde \f^2 + { d U^2 \over 4 \tilde f(U)}
\,, \\ \tilde f(U) &=&  U^2 +\a U + { J \over 2} \label{genBTZ}\,,\\
T_+ &=&  2\p { 1 + r_0^2 \over 1 - r_0^2}\,,\\
G_+ &=& 0\,,\\
A^L_+ &=& W_+ {l \over 2} d \tilde  \f\,,\label{outside}
\eea
where, as before, $\a$ is defined by $J - M = \a^2 $. The parameters in the outside solution are given by
\bea
\a &=& 1- W_+ \label{alphasol}\,,\\
J&=&{\mu W_+^2 \over2 ( \mu-1)} \label{Jsol}\,,\\
M &=& J - (1 - W_+)^2 \label{Msol}\,,\\
U_0 &=& {\m W_+ \over 2(\m -1)} \label{U0sol}\,.
\eea

As a consistency check, it is easy to see that the metric of the complete solution is continuous. However, verifying whether the equations of motion are satisfied, by checking if indeed the jumps in the other fields and derivatives of the metric are exactly those produced by the domain wall source takes a little more work. We detail the calculation in the following subsection.

\subsubsection{The Israel matching conditions}
We now show that the above configuration satisfies the Israel matching conditions \cite{Israel:1966rt,Levi:2009az}, needed to ensure the glued solution satisfies the equations of motion also at the position of the delta-function sources. The Israel matching conditions follow from considering the  bulk action (\ref{20action}) coupled
to a domain wall  action given by
\bea
S &=& {1 \over 16 \p G} \int d^3x \left[ \call_{bulk} + {\d( r - r_0) } \call_{DW} \right] \label{matchaction}\\
\call_{bulk} &=& \sqrt{-g}\left(R+\frac{2}{l^2}-(\m -1)\left( {\pa_\m T \pa^\m T  + \half G_{\m\n}G^{\m\n} \over T^2 - 4 \p^2} \right) \right)
- { 2 \over  l} \e^{\m\n\r}A^L_\m \pa_\n A^L_\r \,,\\
\call_{DW} &=&  {2   \r \over \sqrt{g_{rr}}} \left[ ( (\m -1) T
- 2\p b) \sqrt{ - g_{tt}} - (\m -1) C_t + 2\p b A^L_t \right]\,. \eea
We derive the solution presented in the previous subsection by
explicitly showing how the parameters $(r_0,b,\rho)$ and
$(U_0,J,M)$ are fixed to the values given above by the matching
conditions.


The matching conditions require  the metrics $ds^2_\pm$ to be
continuous across the matching surface. This determines the parameters $U_0, J$ 
\bea
U_0 &=&  {\m r_0^2 \over 1- r_0^2}\,,\\
J &=& {2 \m r_0^2 \over (1-r_0^2)^2}\left( 1 - \m r_0^2 -  \a(1-r_0^2)\right)\,.
\eea


We are now ready to solve the jump conditions on the fields across the shell. First of all, it is easy to see that the Bianchi identity  $dG = 0$ is obeyed
everywhere without delta-function terms. This is because $G_-$ only has legs along the $r$-direction.
Next, we note that the  values of $\r$ and $b$ are fixed by axionic and Chern-Simons charge conservation.
Since the G\"odel background has one unit of $(\p,\p,0)$ axionic charge and the axionic charge on the outside  vanishes, the total $(-\p,-\p,0)$
charge on the wall also has to add up to one.
This fixes $\r$ to be
\be \rho = {1 \over 2 \p } \label{rhosol}\,,\ee
Similarly, the total Chern-Simons charge on the wall is given by $  2 \p  \r b=  b$. On the outside we have charge $W_+$, while on the inside we have $W_-$, so $b $ is
\be
b = W_+ - W_- = W_+ + \m -1\label{bsol}\,.
\ee
Plugging into (\ref{wallpos})  we find that the shell is located at
\be
r_0^2 = {W_+ \over 2(\m -1) + W_+} \label{r0sol}\,.
\ee
Note that, for our embedding in string theory discussed in Appendix \ref{IIBappendix}, where $\m = 3/2$, $b$ becomes
\be b  = \half  + { R \over  N^2}  \,,\ee
and is allowed by the quantization condition from string theory (\ref{bquant}).
In Appendix \ref{appmatch}, we show that these values for $\r,b$ obtained by naive charge conservation indeed solve the precise Isra\"el conditions for the fields $T, C$ and $A^L$.

Now we turn to the matching condition on the metric. Since our brane source is massless by construction, this amounts to requiring continuity of the
extrinsic curvature across the matching surface:
\be
K^+_{ij} = K^-_{ij} \,,\label{gravmatch}
\ee
with
\be  K_{ij} = \half \call_n h_{ij} \,,\ee
where $n$ is the unit normal to the surface.
The extrinsic curvatures are given by
\bea
K^+_{ij} &=&  {l \sqrt{\m} r_0  \over (1 - r_0^2) } \left( \begin{array}{cc} 0 & -  1 \\  - 1 &  \a \end{array}\right)\,,\\
K^-_{ij} &=& {l \sqrt{\m} r_0  \over (1 - r_0^2) } \left( \begin{array}{cc} 0 & - 1 \\  - 1   & {1 - ( 2 \m - 1)r_0^2 \over 1 - r_0^2} \end{array}\right)\,.\\
\eea
The gravitational matching equations (\ref{gravmatch}) then require
\bea
\a &=& { 1 - {r_0^2 \over r_B^2} \over 1- r_0^2} \label{matchsol}\,.
\eea
Here we have defined a special radius
\be
r_B = {1 \over \sqrt{2 \m -1}}\,.
\ee
where $\a$ becomes zero and to which we will refer to as the `Bousso radius' for reasons to be explained in section \ref{extremelimit}. 
Utilizing (\ref{r0sol}) we finally reproduced the complete solution as presented above.

\subsubsection{Global supersymmetry}
Let us now verify that our matched spacetime preserves half of the supersymmetries. The inside
G\"odel solution is half-BPS as shown in section \ref{globalgodel}. In section \ref{susyBTZ} we saw that, for NS sector boundary conditions on the fermions,
the BTZ metric is supersymmetric if $\a$ is related to the Wilson line $W_+$ as
\be
\a = 1- W_+\,.
\ee
Plugging (\ref{r0sol}) into (\ref{matchsol}), we find that
this is indeed precisely the relation obeyed by our outside geometry.
We can also explicitly see that our matched solution preserves global Killing spinors.
These were derived in (\ref{eq:Godel_Spinor}) and (\ref{eq:KS_BTZ}) for the G\"odel and BTZ geometries respectively\footnote{Note that these
expressions were derived using a vielbein that is continuous across the matching surface, so that we can compare them
without having to make an extra local Lorentz transformation.} and are given by
\be
\e_+ = \e_- = e^{ - i t } \e_0 \qquad {\rm with}\ \g_{\hat{\bar z}} \e_0 = 0\,.
\ee
This establishes that our matched solution globally preserves half of the supersymmetries of (2,0) supergravity: it consists of
two  half-BPS solutions, glued across a thin shell  in such a way that the
Killing spinors are continuous, using matter on the shell that is separately supersymmetric.

Putting all this together we obtain the matched solution given in (\ref{inside}-\ref{U0sol}).
It's also useful to look at the quantum numbers of our glued  space in the dual CFT:
\bea
 L_0 &=& {c \over 24} (M-J+1 + W_+^2 )= {c \over 12} W_+  \\
R &=& {c \over 12} W_+\\
 \bar L_0 &=&{c \over 24}( M + J+1) = {c \over 24} {W_+ ( W_+ + 2 (\m - 1))\over \m - 1}
\eea
The first two equations are the characteristic quantum numbers of a chiral primary in the left-moving
superconformal algebra.  The third equation tells us that we have a particular excited state on the right-moving side. Of course, we could modify the value of $\bar L_0$ by  adding
 rightmoving  excitations of extra matter fields, present in the string embedding of our system,  such as rightmoving Wilson lines.

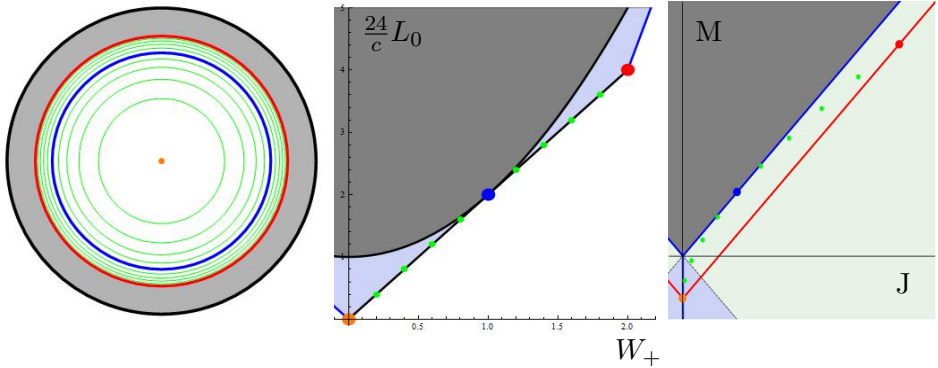
\begin{figure}[t]
\begin{center}
\begin{picture}(300,150)
\put(-20,0){\includegraphics[width=120pt]{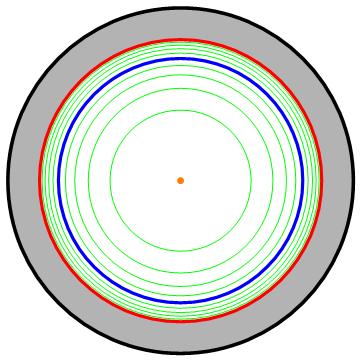}}
\put(105,-5){\includegraphics[width=120pt]{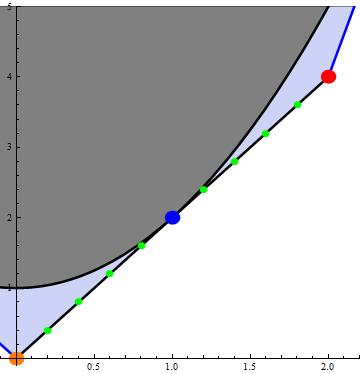}}
\put(115,105){${24 \over c} L_0$} \put(210,-15){$W_+$}
\put(230,0){\includegraphics[width=100pt]{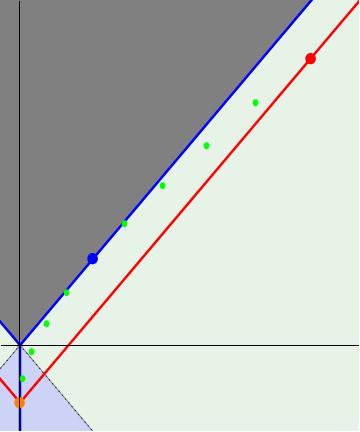}}
\put(240,105){M} \put(315,10){J}
\end{picture}
\end{center}
\caption{The family of matched solutions obtained. Left: position
of the domain wall in G\"odel. Middle: their dual CFT quantum
numbers. Right: their mass and angular momentum. For zero R-charge
(in orange), the G\"odel region has zero size and we simply have
global $AdS_3$. Increasing the R-charge progressively increases
the size of the G\"odel region. The Bousso radius (in blue)
corresponds to the maximal entropy configuration on the black hole
threshold. The maximal radius beyond which  there would be CTCs (in
red) corresponds to the maximum R-charge allowed by unitarity.}\label{wallsfig}
\end{figure}

Summarizing, we have found asymptotically AdS solutions
corresponding to the backreaction of a solid cylinder  of G\"odel
space  with massless RQ particles smeared on the edge of the
cylinder. In the dual CFT, the left moving quantum numbers  are
those of a chiral primary state of the (2,0) superconformal field
theory
 while the rightmovers are in a specific non-supersymmetric excited state.
We have obtained in this way a family of matched solutions that
are characterized by the total R-charge and illustrated in Figure
\ref{wallsfig}.

\subsubsection{Chronology protection and stringy exclusion}

Let's discuss our matched solution (\ref{matchedsol})  from the point of view of chronology protection.  We found that our massless branes condense at radius
\be
r_0^2 =  {W_+ \over 2(\m -1) + W_+}\,.
\ee
We can  vary the position $r_0$ of the domain wall by varying the
R-charge  of the matter on the wall, which will change the value
of $W_+$. From the bulk  gravity point of view, there are
constraints on the value of $r_0$.
The first requirement is  that $r_0$ is positive; this simply  leads to $W_+ \geq 0$ or $L_0\geq0$ as expected from \cite{Raeymaekers:2009ij}.
A second constraint arises when one requires that the glued space is free of closed   timelike curves. For this  we should impose
$r_0^2\leq1/\mu$ which gives gives $W_+ \leq 2$. One easily checks that this also implies that the outside metric
is free of CTCs.
These requirements agree precisely with the  unitarity bounds on
chiral primaries  (\ref{unitboundCP}) in the dual CFT, the latter
bound being the stringy exclusion principle (see Figure \ref{wallsfig}).
Hence our construction gives a bulk  gravity explanation for the
stringy exclusion principle!

Of course one can also turn this argument  around, and observe
that if one imposes the unitarity bound of the dual CFT on the
bulk solutions, that the remaining matched solutions are free of
closed timelike curves. In this sense our result here is a
generalization of the observation made in
\cite{Raeymaekers:2009ij}, that unitarity on the boundary seems
equivalent to chronology protection in the bulk.

\subsubsection{The extremal black hole limit}\label{extremelimit}

There is a special value of $W_+$, namely \be W_+ = 1\,, \ee which
is interesting both from the point of  view of the inside as well
as the outside geometry (see figure \ref{wallsfig}).

From the point of view of the G\"odel space, the gluing radius becomes equal to what we called the Bousso radius:
\be
r_0 = r_B ={1 \over \sqrt{2 \m -1}}\,.
\ee
This radius is special in that it is the radius
where the $\f$ circle has maximum length.
In Bousso's proposal for holographic descriptions of general space-times \cite{Bousso:1999cb}, this is where
 the optimal holographic screen is located \cite{Boyda:2002ba}. Therefore the worldvolume theory on our domain wall
 would be a candidate for a holographic dual description.

In this particular case, the outside metric has $M = J>0$ and lies at the extremal black hole threshold.
As we can see from (\ref{newBTZcoords}), the outside metric is really a scaling limit of the BTZ metric, controlled by the parameter $\a = \sqrt{J-M}\to 0$. In this limit, we approach
extremality  while we take a near-horizon limit  and rescale energies at the same time.
In this case,  the outside metric can be written as
\be
\frac{ds^2_+}{l^2} = - {2 \over J} (U^2 + {J \over 2} ) dt^2 + {d U^2 \over 4 (U^2 + {J \over 2} )} + {J \over 2} ( d\f + {2 U \over J} d t)^2\,.
\ee
Hence we obtain a circle fibration over $AdS_2$. This scaling limit goes under the name of the selfdual orbifold and was considered in \cite{Coussaert:1994tu} (see
\cite{Balasubramanian:2010ys} for a recent discussion).
The limit also corresponds to special quantum numbers in the dual CFT: the corresponding ensemble of BPS states has maximal degeneracy within the family of chiral primaries.

The degeneracy of chiral primaries has a symmetry around this
middle value of the R-charge. As we can think of the chiral
primaries as represented by harmonic forms, with their degree
corresponding to the R-charge \cite{Maldacena:1998bw}, Hodge
duality then indeed provides a map between states of charge $R$
and those of charge $R_{\mathrm{max}}-R$. In our notation, as
specified in table \ref{tab:StringyExclus}, this corresponds to
$W_+\rightarrow 2-W_+$.  It is interesting to note that this
symmetry is reflected in the matched solutions we described above.
If one computes the length of the circle on which the domain wall
is smeared one finds: \be L=2\pi
l\sqrt{\frac{\mu}{4(\mu-1)}W_+(2-W_+)}\,, \ee which is indeed
invariant under this 'reflection of R-charge'. This is suggestive
of a relation between  the degeneracies of the chiral primaries
and the area of the domain wall, it would be interesting to attempt
a quantization of the domain wall and investigate what degeneracies
one finds and how they relate to that of the chiral primaries in
the theory.

\subsubsection{Spectral flow}
The solutions discussed above obey NS boundary conditions on the spinors.
 We can obtain configurations with different boundary conditions on the spinors by performing an integral spectral
 flow on the matched solution, which corresponds in the bulk to shifting both Wilson lines $W_+, W_-$ with an integer number $n$. The resulting configuration
 is
 \bea
J&=&{\mu (W_+ + n)^2 \over(2 ( \mu-1))} \,,\\
M &=& J - (1-n - W_+)^2\,,\\
r_0^2 &=& {W_+ + n \over 2(\m -1) + W_+ + n} \,,\\
U_0 &=& {\m ( W_+ + n)\over 2(\m -1)}\,,\\
b &=&  W_+ + \m -1\label{matchedsol} \,.
\eea
The preserved Killing  spinors are then
\be
\e_+ = \e_- = e^{-it + i{n \over 2} \f} \e_0 \qquad \mbox{with}\   \g_{\hat{\bar z}} \e_0 = 0 \qquad (\Leftrightarrow \g^{\hat t} \e_0 =  i \e_0)\,.
\ee

A  discussion similar to the one above  gives the limits
\be
-n \leq W_+ \leq -n+2\,.
\ee
which again corresponds to the unitarity bounds in these sectors.
For example, at $n = 1$, the outside solutions are R sector ground states, and for $n = 2$ we have antichiral primaries.

\section{Conclusions and outlook}\label{secconc}
In this paper, we proposed a stringy  mechanism for excising the
closed timelike curves in the three-dimensional G\"odel universe.
We started by constructing explicit solutions arising from the
backreaction of 7-branes wrapping an internal $S^3 \times T^4$,
and showed that these produce quotients of the G\"odel metric. A
special role was reserved for the brane with charges $(\p,\p,0)$
which produces the global G\"odel metric accompanied by a Wilson
line.

Next we studied probe branes in this background and identified certain branes that become light in some region of the space. We included these states
in the effective description by assuming they condense on a thin shell at the radius where they become massless. We then found the geometry on the outside
of the shell by solving the Israel matching conditions, and showed that the construction is supersymmetric. We found in this way a discrete one-parameter family of such glued configurations, labeled
by their asymptotic R-charge as illustrated in figure \ref{wallsfig}. Our main observation was that demanding the absence of closed timelike curves in this family of solutions
is equivalent to imposing the stringy exclusion bounds on chiral primaries in the dual CFT.
We see this as evidence for the consistency of our proposal.

Our analysis has revealed some suggestive features that we feel deserve further study.
One such feature is the apparent  correspondence between the length of circle on which branes condense  and the microscopic degeneracy of the outside configuration viewed as an ensemble in the dual CFT. In particular, the symmetry of the degeneracies appears to
be correspond to  a symmetry of the area under a reflection around the maximal  area curve. The maximal length curve corresponds to to the location
of Bousso's  optimal holographic screen, and the corresponding CFT quantum numbers are those of the maximal entropy configuration on the
black threshold. These observations point to the possibility that the worldvolume theory of the matter on the shell would be a natural candidate for a holographic description of our glued spacetime. The model studied here could in this sense serve as  a test case for understanding holography in more general non-asymptotically AdS spaces,
 such as Kerr-CFT \cite{Guica:2008mu}, warped AdS backgrounds \cite{Anninos:2008fx} or de Sitter space \cite{Strominger:2001pn}.
To make progress along these lines one would need a better understanding of the worldvolume degrees of freedom of our condensing branes.
We saw that they have an internal angular momentum component and a generalized 7-brane component; it would be interesting
to see if the thin shell we constructed can be seen as a type of giant graviton hypertube.

Finally, let us comment on the relation  to \cite{Levi:2009az},
where supersymmetric G\"odel embeddings were found in M-theory,
corresponding to the backreaction of wrapped M2-branes. Treated as
probes these M2 branes were shown to provide the correct degrees
of freedom to describe the entropy of the D4D0 black hole
\cite{Denef:2007yt}. One would now like to apply the new insights
obtained in the present paper, namely how the addition of
Q7-branes can resolve the closed timelike curves of the G\"odel
space, to that setup, by going through a chain of U-dualities
\cite{Levi:2009az}. It would be interesting to see if this could
lead to a more precise description of the deconstruction model
\cite{Denef:2007yt} that takes backreaction into account, and if
the first hints we found concerning the degeneracies of chiral
primaries and G\"odel holography could be made more concrete in
that setting.

\section*{Acknowledgements}
The authors are pleased to acknowledge useful discussions with
I.~Bena, F.~Denef, A.~Royston and W.~Troost.

The work of JR  has been supported by the EURYI grant GACR  EYI/07/E010 from EUROHORC and ESF. DVdB is supported by the DOE under grant DE-FG02-96ER4094 and he thanks the Feza G\"ursey Institute, Bo\u{g}azi\c{c}i University  and CEA-Saclay for their hospitality while part of this work was completed. The work of B.V. was supported in part by the Federal Office for
Scientic, Technical and Cultural Affairs through the
Interuniversity Attraction Poles Programme Belgian Science Policy
P6/11-P, by the FWO-Vlaanderen through both an aspirant fellowship
and the project G.0235.05 and by DSM CEA-Saclay, by the ANR grant
08-JCJC-0001-0, and by the ERC Starting Independent Researcher
Grant 240210 - String-QCD-BH.

\begin{appendix}

\section{Type IIB embedding of (2,0) supergravity}\label{IIBappendix}
In this appendix we show the reduction and consistent truncation that embeds (2,0)-supergravity (\ref{20action}) into Type IIB supergravity. We show this for the equations of motion, the generic 7-brane actions of \cite{Bergshoeff:2006jj} and the supersymmetry variations of the theory.

\subsection{Type IIB conventions}
We start from type IIB supergravity in the formalism of \cite{Schwarz:1983qr}, truncated to include gravity, the
selfdual five-form and the axion-dilaton. We define SL(2,$\mathbb{R}$)/SO(2) coset 1-forms
$\cP,\kQ$ given in terms of the standard axiodilaton $\T$ as \bea
\cP &\equiv&  { d \T \over \sqrt{2} \mathrm{Im}\,\tau}\,,\\
\kQ &\equiv&  {1 \over \mathrm{Im}\T} d\mathrm{Re}\T,.
\eea
They satisfy
\bea
d \kQ =  i \cP \wedge \bar \cP\,,\\
d \cP + i \kQ\wedge \cP = 0\,.
\eea
The (pseudo-) action is
\be S = {2 \p \over
l_s^8} \int_{\calm_{10}} \left[ \star R -  \cP \wedge \star \bar \cP - {1 \over2 \cdot 5!}  F_5 \wedge \star F_5
\right]\,,\ee
where, as usual, we have to impose the selfduality
constraint $F_5 = \star F_5$ at the level of the equations of
motion.

The equations of motion are
\bea
R_{MN} &=&{1 \over 2}( \cP_M \bar \cP_N +  \cP_N \bar \cP_M) + { 1\over 4 \cdot 4!} F_{MPQRS}F_N^{\ PQRS}\,,\\
(\nabla^M + i \kQ^M) \cP_M &=&0\,,\\
F_5&=& \star F_5\,.
\eea

We will also consider Q7-brane sources \cite{Bergshoeff:2006jj,Bergshoeff:2007aa}. These couple to the
supergravity fields as \be S_{Q} = {2 \p \over l_s^8} \left[
-\int_{\calm_8} T_{Q} \sqrt{ - g} + \int_{\calm_8} C_{8\, Q}\right] \label{branes10d}\,.
\ee
Here $Q$ is a traceless real matrix, parameterized by three charges, ($p,q,r$),
\be
Q=\begin{pmatrix}\frac{r}{2}&p\\-q&-\frac{r}{2}\end{pmatrix}\,.
\ee
The tension and the precise 8-form that the 7-brane couples to depend on $Q$:
\bea
T_Q&=&\frac{1}{\mathrm{Im}\,\tau}\left(p+r\mathrm{Re}\tau+q|\tau|^2\right)=2\sqrt{\det Q}\coth\left(\sqrt{\det Q}\mathrm{Im}\,\sw_Q\right)\,,\\
dC_{8\, Q}&=&\frac{\sqrt{\det Q}\star \!d \mathrm{Re}\, \sw_Q}{\sinh^2\left(\sqrt{\det Q}\mathrm{Im}\,\sw_Q\right)}\,.
\eea
The relation between $\tau$ and $\sw_Q$ is given by
\be
\begin{cases}
\tau=-\frac{r}{2q}-\frac{\sqrt{\det Q}}{q}\cot\left(\sqrt{\det Q}\,\sw_Q\right)\qquad\mbox{when }q>0\\
\tau=\frac{p}{r}\left(e^{r\sw_Q}-1\right)\ \ \,\quad\qquad\qquad\qquad\qquad\mbox{when }q=0\,.
\end{cases}
\ee

\subsection{Reduction ansatz}
We are interested in compactifications of the above theory
on $S^3 \times T^4$,
with  Q7-brane sources wrapping on $S^3 \times T^4$. For this,
we take the following ansatz for the fields: \bea
ds^2 &=& ds^2_3 + l^2 d\tilde\O_3^2 + V^\half ds^2_{T^4}\,,\\
d\tilde\O_3^2 &=& {1\over 4} \left[ d\theta^2 + \sin^2\theta d\psi_R^2  + ( D_L \psi_L + \cos\theta d\psi_R)^2 \right]; \qquad D_L \psi_L \equiv d \psi_L + {4 \over l} A^L\label{covder}\,,\\
F_5 &=& {2 \over l}  (e^0\wedge e^1\wedge e^2+ e^3\wedge e^4\wedge e^5 ) \wedge S_+\,, \\
\T &=& \T (x^a) \label{redansatz}\,,
\eea
where $x^a$ are the coordinates on the remaining 3d manifold, $ds^2_{4}$ is the unit volume flat metric on $T^4$ and $S_+$ is a closed, selfdual 2-form on  $T^4$: $d S_+ =0,\ S_+ = \star_4 S_+$.
We have allowed a nontrivial fibration of the $S^3$ by a 3D gauge field $A$; this will give rise to a $U(1)$ Chern-Simons field in 3D. This $U(1)$ sits in the $SU(2)_L$ of the
$SU(2)_L \times SU(2)_R$ of the sphere isometry group; by interchanging the $L$ and $R$ labels one gets a $U(1)$ in $SU(2)_R$. The Bianchi identity for $F_5$ requires that $F^L \equiv d A^L =0$.

In general we can decompose
\be
S_+ = n^i s_i, \qquad n \cdot n = 1\,,
\ee
with
\bea
s_1 &=& {1\over \sqrt{2}} (e^6 \wedge e^7 + e^8 \wedge e^9 )\,,\\
s_2 &=& {1\over \sqrt{2}} (e^6 \wedge e^8 + e^9 \wedge e^7 )\,,\\
s_3 &=& {1\over \sqrt{2}} (e^6 \wedge e^9 + e^7 \wedge e^8 )\,,
\eea
so that $\int_{T^4} S_+ \wedge S_+  = V$.  The $S_i$ transform as a triplet under the $SU(2)_L$ subgroup of the $SO(4)$ frame rotations;  we will later use this freedom to
fix $n^i$ to a convenient value without loss of generality.
Flux quantization tells us that the $F_5$ flux through any cycle should be an integer multiple of $l_s^4$; this gives the quantization condition
\be
l^2 = {N l_s^4 \over 4 \p^2 V^\half}\,.
\ee

Verifying the equations of motion, one finds that the $S^3$ and $T^4$ parts are satisfied\footnote{It's easiest to use an orthonormal basis to check this.} and one is left with the 3D equations
\bea
R_{ab} + {2 \over l^2}g_{ab}&=& \cP_a \bar \cP_b +  \cP_b \bar \cP_a\,,\\
(\nabla^a + i \kQ^a) \cP_a &=&0\,,\\
d F_L &=& 0\,.
\eea
These can be derived from an effective 3D action
describing $AdS$ gravity coupled to an axion-dilaton system and a $U(1)$ Chern-Simons gauge field:
\be S = {2 \p \over
l_3} \int_{\calm_{3}} \left[ \star (R + {2\over l^2})- 2  \cP \wedge \star \bar \cP - {1 \over l} A^L \wedge F^L
\right]\,. \ee
 This is precisely the (2,0) action (\ref{20action}) with the parameter $\m$ fixed to $\mu = 3/2$. The three-dimensional Planck length is
\be
l_3 = {l_s^8 \over 2 \p^2 l^3 V}=32\pi^2 G\,.
\ee
From these relations we find the Brown-Henneaux central charge:
\be
c = {3 l \over 2 G} = { 48 \p^2 l \over l_3} = 6 N^2\,.
\ee
The U(1) Chern-Simons field is part of an SU(2)$_L$ Chern-Simons field. Since $\p_3 (SU(2)) = \mathbb{Z}$,   the
coefficient of the  action is quantized in terms of the integer level $k$ 
 In our normalization, the level is
\be
k =  { 8 \p^2 l \over l_3}= {c \over 6}   = N^2\,,
\ee
which is indeed integer.

\subsection{BPS source terms}
Here we discuss how the 3d point particle actions discussed in section \ref{susypart} follow from reduction.

First look at the 3D source terms produced by a Q7-brane wrapped on $S^3 \times T^4$.
Reducing (\ref{branes10d}) one finds \be S_{Q} = {1 \over 16 \p
G}\left[ - \int T_{Q} ds + \int C_Q\right]\,. \ee We will also
consider consider another BPS source, namely a BPS particle
charged under $A$. Is this a KK mode on the three-sphere. Such
source terms produce the BPS conical defects discussed in
\cite{Balasubramanian:2000rt}.  The action is \be S_{CD} = {b\over
4 G}\left[ - \int  ds +  \int A^L \right]\,, \ee where $b$ is the
mass (= charge) of the particle.

Let's find the quantization condition on $b$. In the dual CFT, the
$U(1)_L$ R-charge $R$ is an integer. It corresponds to an
asymptotic Wilson line going as \be A^L = {R \over k} l d\f\,. \ee
The field strength has a delta-singularity: \be F^L = {R \over k}
l \d ( r) dr\wedge d\f\,. \ee This in turn is produced by a source
term \be { R \over  l}\int A^L\,. \ee Hence $b$ is quantized as
\be b = { R \over k} = { R \over N^2} \label{bquant}\,. \ee

\subsection{Spinor reduction}
We now discuss the reduction ansatz for the spinors. We will first reduce to six dimensions and then use the results
of \cite{Balasubramanian:2000rt} to further reduce to 3D.
The type IIB Killing spinor equations coming from the gravitino and dilatino variations are
\bea
\left( \nabla_M - {i \over 4}\kQ_M + { i \over 8 \cdot 4!} F_{MNPQR}\G^{NPQR} \right) \e_{10} &=& 0\\
\G^M \cP_M \e^c_{10} &=& 0\,.\label{10DKS}
\eea
Here, $\e_{10}$ is a 10D Weyl spinor
\be
\G_{11} \e_{10} = \e_{10}, \qquad \G_{11} \equiv \G^0 \G^1 \ldots \G^9
\ee
and conjugation is defined as
\be
\e^c_{10} = (B_{10} \e_{10})^*, \qquad B_{10} \G^M B_{10}^{-1} = \G^{M*},\ B_{10}^* B_{10} =1\,.\label{10Dconj}
\ee
Using these properties, the complex conjugates of (\ref{10DKS}) are equivalent to
\bea
\left( \nabla_M + {i \over 2}\kQ_M - { i \over 8 \cdot 4!} F_{MNPQR}\G^{NPQR} \right) \e_{10}^c &=& 0\,,\\
\G^M \bar \cP_M \e_{10} &=& 0\,.\label{10DKSconj}
\eea

We decompose the gamma-matrices according into $SO(1,5) \times SO(4) \subset SO(1,9)$
\bea
\G^\m &=& \tilde \g^\m \otimes \g_{(5)}\qquad \m = 0,\ldots, 5\\
\G^i &=& 1 \otimes \g^i\qquad i = 6,\ldots, 9\\
\G_{(11)} &=& \g_{(7)} \otimes \g_{(5)}\\
B_{10} &=& B_6 \otimes B_4\\
\g_{(7)} &=& \tilde \g^0 \tilde \g^1 \ldots \tilde \g^5\qquad \g_{(5)} = \g^6\g^7 \g^8 \g^9\\
B_6  \tilde \g^\m B_6^{-1} &=&  \tilde \g^{\m*}, \qquad B_6^* B_6 = -1\\
B_4 \g^i B_4^{-1} &=& \g^{i*}, \qquad B_4^* B_4 = -1\,.
\eea
Now we expand in a special basis of $SO(4)$ spinors. Because the Weyl spinor representation of $SO(4)$ is pseudoreal, we can find basis spinors $\h^{\pm}_A,\ A=1,2$
that satisfy
\bea
\g_{(5)} \h^\pm_A &=& \pm \h^\pm_A\,, \\
(B_4 \h^\pm_A)^* &=& \e^{AB} \h^\pm_B\qquad \e^{12} \equiv 1\,.\label{basisprop}
 \eea
We now decompose the 10D spinors $\e_{10},\ \e_{10}^c$ into 6D spinors $\Psi^{\pm }_{l}, \ l = 1,\ldots ,4$ according to the ansatz
\bea
\e_{10} &=& \Psi^+_1 \otimes \h^+_1  +\Psi^+_4 \otimes \h^+_2 +\Psi^-_1 \otimes \h^-_1 +\Psi^-_4 \otimes \h^-_2 \,,\\
\e_{10}^c &=& \Psi^+_2 \otimes \h^+_1  +\Psi^+_3 \otimes \h^+_2 +\Psi^-_2 \otimes \h^-_1 +\Psi^-_3 \otimes \h^-_2\,.
\label{spinorred}
\eea
The 10D chirality and reality condition (\ref{10Dconj}) reduce to conditions on the 6D spinors:
\bea
\g_{(7)} \Psi^\pm_m &=& \pm \Psi^\pm_m \qquad m = 1,\ldots,4\,,\\
( B_6 \Psi^\pm_m )^* &=& \O_m^{\ n}\Psi^{\pm}_n\label{symplmaj}\,,
\eea
where
\be
\O_m^{\ n}  =  \left( \begin{array}{cccc} 0 &0&1&0\\0 &0&0&1\\-1 &0&0&0\\0 &-1&0&0 \end{array}\right)\,.
\ee
The second condition is the symplectic Majorana condition, so we have decomposed the 10D spinor into two 6D symplectic Majorana-Weyl
spinors of opposite chirality.

Now we can plug our spinor decomposition ansatz (\ref{spinorred}) into the Killing spinor equations (\ref{10DKS}) using the reduction ansatz (\ref{redansatz}).
Let's start with the $T^4$ components of the gravitino equation. These reduce to
\be
(1 + \g_{(7)} \otimes 1)\e_{10} =0\,,
\ee
or, in terms of the 6D spinors:
\be
\Psi^+_m = 0\,.
\ee
so we are left with a single symplectic Majorana-Weyl spinor $\Psi^-_m$.

Now we write out the susy conditions on the 6D spinor $\Psi^-_m$. We define the 6D selfdual 3-form $H$ as
\be
H_3 = {1 \over l} (e^0\wedge e^1\wedge e^2+ e^3\wedge e^4\wedge e^5 )\qquad *_6 H_3 = H_3\,.
\ee
The operator $S_{+ij}\g^{ij}$ generates an $SU(2)_L \subset SO(4)$ rotation under which the $\h^-_A$ transform as a doublet and the $\h^+_A$ as a singlet. By choosing a suitable frame
on $T^4$ we can arrange that
\be
\half S_{+ij}\g^{ij}\h^-_A =  i \s^{3\ B}_A \h^-_B.
\ee
Then the 6D part of the KS equations (\ref{10DKS},\ref{10DKSconj}) reduces to
\bea
\left( \nabla_\m - {i \over 2}\kQ_\m - { 1 \over 4} H_{\m\n\r} \tilde  \g^{\n\r} \right) \Psi^-_1 &=& 0\,,\\
\left( \nabla_\m + {i \over 2}\kQ_\m + { 1 \over 4} H_{\m\n\r} \tilde \g^{\n\r} \right) \Psi^-_2 &=& 0\,,\\
\left( \nabla_\m + {i \over 2}\kQ_\m - { 1 \over 4} H_{\m\n\r} \tilde \g^{\n\r} \right) \Psi^-_3 &=& 0\,,\\
\left( \nabla_\m - {i \over 2}\kQ_\m + { 1 \over 4} H_{\m\n\r} \tilde \g^{\n\r} \right) \Psi^-_4 &=& 0\,,\\
\g^\m \cP_\m \Psi^-_1 &=& 0\,,\\
\g^\m\bar \cP_\m \Psi^-_2 &=& 0\,,\\
\g^\m \bar \cP_\m \Psi^-_3 &=& 0\,,\\
\g^\m \cP_\m \Psi^-_4 &=& 0\,.
\eea
For $\cP = \kQ = 0$, we recover precisely the equations (40) in \cite{Balasubramanian:2000rt}.
Although we wont do this explicitly, let us point out that these eqs. should be embeddable in 6d N =4b supergravity with tensor multiplets.

Next we reduce to 3D using the conventions of \cite{Balasubramanian:2000rt}. We decompose the Clifford matrices as
\bea
\tilde \g^\a &=& \s_1 \otimes 1 \otimes \g^\a \qquad \a = 0,1,2\,,\\
\tilde \g^a &=& \s_2 \otimes \g^a \otimes 1\qquad a = 3,4,5\,,\\
\g^0 &=& - i \s_2 \qquad \g^1 = \s_1 \qquad \g^2 = \s_3\,,\\
\g^3 &=&  \s_1 \qquad \g^4 = \s_2 \qquad \g^5 = \s_3\,,\\
\g_{(7)} &=& \s_3 \otimes 1\otimes 1\,,\\
B_6 &=& \g^3\g^5 = 1 \otimes -i\s_2 \otimes 1\,.
\eea
The only difference with the conventions of \cite{Balasubramanian:2000rt} is that we have chosen another $B_6$. We will need the useful identities
\bea
\g^{\a\b} &=&  \e^{\a\b\d}\g_\d\,,\\
\g^{ab} &=&  i\e^{abc}\g_c\,.
\eea
The chirality condition means that we can write
\be
\Psi^-_m =  \left( \begin{array}{c} 0 \\1  \end{array}\right) \otimes \varepsilon_l\,,
\ee
with $\varepsilon_l$ 4-component spinors. Now we work out the $S^3$ part of the gravitino equation. For this it is convenient to choose
the left-invariant vielbein:
\bea
e^3 &=& {l \over 2} ( \cos \psi_L d \theta + \sin \psi_L \sin \theta d\psi_R )\,,\\
e^4 &=& {l \over 2} ( -\sin \psi_L d \theta + \cos \psi_L \sin \theta d\psi_R )\,,\\
e^5 &=& {l \over 2} ( D_L \psi_L  + \cos \theta d \psi_R )\,.
\eea
The spin connection is
\bea
\o_{34} &=& - {1\over l} (e^5 - 2 A^L )\,,\\
\o_{53} &=& - {1\over l} e^4 \,,\\
\o_{45} &=& - {1\over l} e^3\,.
\eea
If we take the Killing spinor to be independent of the $S^3$ coordinates, the $S^3$ gravitino equation projects
\be
\varepsilon_1 = \varepsilon_3 =0.
\ee
The remaining 3D equations for $\varepsilon_{2,4}$ are
\bea
\left( \pa_\a + {1\over 4}\o_{\a}^{\hat \b \hat \g} \e_{\hat \b \hat \g \hat \d} 1 \otimes \g^{\hat \d} - { 1 \over 2 l}e_\a^{\hat\a} 1\otimes\g_{\hat \a} + {i \over l} A^L_{\a} \s_3 \otimes 1 + {i\over 4}\kQ_\a \right)
\varepsilon_2 &=& 0 \,,\\
\left( \pa_\a + {1\over 4}\o_{\a}^{\hat \b \hat \g} \e_{\hat \b \hat \g \hat \d} 1 \otimes \g^{\hat \d} - { 1 \over 2 l}e_\a^{\hat\a} 1\otimes\g_{\hat \a} + {i \over l} A^L_\a \s_3 \otimes 1 - {i\over 4}\kQ_\a \right)
\varepsilon_4 &=& 0 \,,\\
\bar \cP_\a 1 \otimes \g^\a \varepsilon_2 \ &=& 0\,,\\
\cP_\a 1 \otimes \g^\a \varepsilon_4 \ &=& 0\label{3Dsusy}\,,
\eea
while the reality condition (\ref{symplmaj}) imposes
\be
(i \s_2\otimes 1) \varepsilon_2^* = \varepsilon_4\,.
\ee

We can now embed the Izquierdo-Townsend solutions \cite{Izquierdo:1994jz}. Suppose we have a 2-component spinor $\e$ solving the equations (\ref{20susy}):
\bea
\left( \pa_\a + {1\over 4}\o_{\a}^{\hat \b \hat \g} \e_{\hat \b \hat \g \hat \d}\g^{\hat \d} - { 1 \over 2 l}e_\a^{\hat\a}\g_{\hat \a} - {i \over l} A^{L}_{ \a} - {i\over 4}\kQ_\a \right)\e &=& 0 \,,\\
\cP_\a \g^\a \e\ &=& 0\,.
\eea
Then we can solve (\ref{3Dsusy}) by embedding it as
\be
\varepsilon_4 = \left( \begin{array}{c} 0 \\ \e  \end{array}\right) \qquad \varepsilon_2 = \left( \begin{array}{c} - \e^* \\ 0   \end{array}\right)\,.
\ee
If the background has $\cP = \kQ =0 $, the number of 10D supersymmetries is doubled, because there is then a second embedding
\be
\varepsilon_4 = \left( \begin{array}{c} \e^* \\ 0 \end{array}\right) \qquad \varepsilon_2 = \left( \begin{array}{c} 0 \\  \e   \end{array}\right)\,.
\ee
Similarly if  $A^L  =0 $ there is  a second embedding
\be
\varepsilon_4 = \left( \begin{array}{c} \e \\ 0 \end{array}\right) \qquad \varepsilon_2 = \left( \begin{array}{c}0\\ \e^*   \end{array}\right)\,.
\ee

As explained in \cite{Balasubramanian:2000rt}, there is a further doubling of 10d supersymmetries if there is also a right moving Wilson line $A^R$ turned on with $A^R = A^L$.
Then by using a right-invariant vielbein on $S^3$ we can find additional embeddings as above with $\varepsilon_4$ replaced by  $\varepsilon_1$ and $\varepsilon_2$
by $\varepsilon_3$. In this way, the maximally susy $AdS_3$ vacuum  of (2,0) supergravity lifts to a IIB solution preserving 16 supersymmetries, while the half-BPS G\"odel and conical defect
solutions of (2,0) supergravity lift to solutions preserving 8 supersymmetries.

\section{Supersymmetry details}\label{app:susy}

We discuss two supersymmetric solutions to the (2,0) supergravity system \eqref{20action}: one where the metric is of BTZ-type and the axidilaton is constant, and another where the axidilaton is holomorphic and the metric is G\"odel space. In both cases we allow a Wilson line for the CS gauge field.

\subsection{BPS equations and notation}

We repeat the Killing spinor equations \eqref{20susy} for convenience:
\bea
\left( \pa_\a + B^L_{\a\hat \b}\g^{\hat \b}  - {i \over l} A^L_\a - {i\over 2}(\m-1)\;^*\kQ_\a \right)\e &=& 0 \,,\label{eq:Grav_Eq}\\
(\m-1)\;^*\cP_\a \g^\a \e\ &=& 0\,.\label{proj}
\eea
with the left gravitational Chern-Simons connection $B^L$:
\be
B^L_{\a\hat \b} = {1\over 4}\o_{\a}^{\ \hat \g \hat \d} \e_{\hat \g \hat \d \hat \b} - { 1 \over 2 l}e_{\a \hat\b}\,.\label{Bdef_app}
\ee

Both the BTZ and the G\"odel metric can be written as a time fibration over Euclidean AdS$_2$. It will simplify the analysis if we choose a vielbein adapted to this fact. Both geometries  can be written as
\be
ds^2 = - (e^{\hat t})^2 + e^{\hat z}e^{\hat{\bar z}} \,.
\ee
We make use of a vielbein of the form:
\bea
e^{\hat t} &=& l(dt +\cJ)\,,\nonu
e^{\hat z} &=& l e^\varphi d z\,,\nonu
e^{\bar{\hat z}} &=& l e^\varphi d \bar z\,.\label{eq:App_Vielbein}
\eea
For the case we consider, the vielbein furthermore obeys the properties:\footnote{The first equation is a rewriting of the Einstein equation \eqref{eq:J_EOM} for the one-form $\cJ$, while the form of the second equation follows from the general solution for $\cJ$ from the Einstein equations \cite{Levi:2009az}, which locally takes the form $\cJ = 2 {\rm Im} (\partial \varphi + (1-\n)\partial \ln \tau_2)$ up to a closed form ($\partial$ denotes a Dolbeaut operator).}
\bea
d e^{\hat t} &=& \frac i l e^{\hat z} \wedge e^{\hat{\bar z}} \,,\label{fibrconds1}\\
d e^{\hat z} &=& -{ 2 i \over l\n} (e^{\hat t}-l d \Lambda)\wedge e^{\hat z}\label{fibrconds2}\,,
\eea
where $\n$ is a constant and $d \Lambda$ a closed one-form on the full three dimensional spacetime.
Using these properties we can compute the spin connection
\bea
\o^{\hat t \hat z  } &=&- {i \over l} e^{\hat z}\,,\nonu
\o^{\hat z \hat{\bar z}} &=& - \frac{2 i}{\nu l} \left( (\n - 2)e^{\hat t} + 2 l d \L\right)\,,
\eea
and the connection $B^L_{\a\hat \b}$:
\be
B^L \equiv B^L_{\a \hat \b} dx^a \otimes e^{\hat \b} = \frac1{l\n} \left( (\n-1)e^{\hat t} + l d\Lambda\right)\otimes e^{\hat t}\,.\label{B}
\ee

\subsection{Supersymmetry for global G\"odel solution}
Our (2,0) supergravity model also has a solution where the axidilaton is holomorphic and the spacetime is globally G\"odel.
The global G\"odel solution corresponds to
\be
\det Q = \pi^2.
\ee
and has the following metric, scalars and Wilson line:
\bea
ds^2 &=& -(dt + \cJ)^2 + e^{2\varphi} dz d\bar z\,,\nonu
A^L &=& \frac l2(\mu -1) d \arg z\,,\nonu
\tau &=& i \frac{1+z}{1-z}\,,\label{eq:App_GodelSol}
\eea
with
\be
\cJ = \mu \frac{|z|^2}{1 - |z|^2} d \arg z\,,\qquad e^{2\varphi} = \frac{\mu}{(1-|z|^2)^2}\,.
\ee
Because of holomorphy of $\tau$, the dilatino equation becomes a projection condition
\be
\g_{\hat{\bar z}} \e = 0 \qquad (\Leftrightarrow \g^{\hat t} \e =  i \e )\label{proj}.
\ee
One easily checks the vielbein choice \eqref{eq:App_Vielbein} satisfies the properties (\ref{fibrconds2}) with
\bea
\n &=& \m\nonu
\L &=& t .
\eea
The relevant quantities entering in the covariant derivative are (see \eqref{Qtau})
\bea
B_L &=& ( dt + { \m -1 \over \m} \cJ) \otimes e^{\hat t}\nonu
A^L &=& (1 - \m ) {l \over 2} d \arg z \nonu
\calq &=& d \arg z + { 2 \over \m} \cJ
\eea
Note that, using the spinor projection condition (\ref{proj}), the nontrivial holonomies nicely cancel between the different connections.
The Killing spinor equations are solved by
\be
\e = e^{-it} \e_0 \qquad \text{with}\   \g_{\hat{\bar z}} \e_0 = 0 \qquad (\Leftrightarrow \g^{\hat t} \e_0 =  i \e_0).\label{eq:Godel_Spinor}
\ee
We conclude that the three-dimensional solution \ref{eq:App_GodelSol} is a 1/2 BPS to (2,0) supergravity.

\subsection{Supersymmetry for BTZ type solution}
We start from the BTZ type metric in the form \eqref{eq:BTZmetric_newcoords} used for the matching of the field equations, with Wilson line $A^L$ also given in (\ref{eq:BTZmetric_newcoords}). First, we rewrite the metric \eqref{eq:BTZmetric_newcoords}, with spatial coordinates $U,\tilde \phi$  in terms of the complex coordinate
\bea
z = \rho e^{i\tilde \phi}\,,
\eea
where $\rho = |z|$ is related to the radial coordinate $U$ as
\be
d \ln \rho = \frac{d U}{2\tilde f(U)}\,.\label{eq:Uphi_z}
\ee
The metric is then given as:
\bea
\frac{ds^2}{l^2} &=& \left(-(dt + \cJ)^2 + e^{2\varphi} dz d \bar z \right)\,,
\eea
where $\cJ$ and $\varphi$ read in terms of $z$:
\bea
\cJ &=& U d\arg z\,,\nonu
e^{2\varphi} &=& \frac{\tilde f(U)}{|z|^2}\,,
\eea
and $U$ is to be seen as a function of $|z|$ through \eqref{eq:Uphi_z}.

It is straightforward to check that the choice of vielbein \eqref{eq:App_Vielbein} satisfies the conditions (\ref{fibrconds1},\ref{fibrconds2}), with
\bea
\n &=& 1\,,\\
\Lambda &=& t + \frac{1-\alpha}2 \arg z\,.
\eea
The gauge connections entering in the spinor covariant derivative are:
\bea
B_L &=& ( dt +{ 1-\alpha \over 2} d\f )\otimes e^{\hat t}\,,\\
A^L &=& \frac l 2 W_+ d\arg z\,,
\eea
Since the complex scalar $\tau = cst$ in this solution, only the gravitino equation is non-trivial. It reads:
\be
\left(\partial_\alpha + (\pa_\alpha t) \g^{\hat t} + \frac 12 \pa_\alpha (\arg z)\big{(}(1-\alpha) \g^{\hat t} - i W_+ \big{)} \right)\e = 0\,.
\ee
We search for spinors with projection condition $\gamma^{\hat t}\epsilon = i \epsilon$ as in the G\"odel solution, eq.\ \eqref{eq:Godel_Spinor}. In order for the spinors to obey sensible boundary conditions (periodic or anti-periodic in $\phi$), we require
\be
W_+ = 1-\alpha - n\,, \qquad n \in \mathbb{Z}\,.\label{eq:W+_condition}
\ee
The solution for the Killing spinor equation then reads:
\be
\epsilon = e^{-it - i\frac n2\phi} \e_0\,, \qquad\text{with}\quad  \g_{\hat{\bar z}} \e_0 = 0 \qquad (\Leftrightarrow \g^{\hat t} \e_0 =  i \e_0) \,.\label{eq:KS_BTZ}
\ee
Hence the BTZ-type solution with Chern-Simons Wilson line obeying \eqref{eq:W+_condition} is a half BPS solution.

Note that the Killing spinors \eqref{eq:KS_BTZ} of the BTZ spaces match to those of the G\"odel solution \eqref{eq:Godel_Spinor} only when $n=0$. This corresponds to having NS boundary conditions for the spinors. When $n\neq 0$, we can shift $W_+ \to W_+ - n$ in order to preserve the G\"odel Killing spinor. This corresponds to performing a spectral flow in the dual CFT from another sector back to the NS sector.

\section{Matching conditions for the non-gravitational fields}\label{appmatch}
In the main text we fixed the constants $\rho$ and $b$ by a quick argument based on charge conservation. Here we show that this physical intuition is correct and that these solutions solve the equations of motion in the presence of the domain wall. Thinking of the domain wall as a $\delta$-function source there should be jumps present in the radial dependence of the fields to cancel this term in the equations of motion. More precisely, the equations of motion at the position of the domain wall take the form:
\be
\Delta I= j\label{matcheq}
\ee
with $\Delta f= f_+-f_-$. Taking $n$ to be a unit normal vector to the matching surface, and  assuming a two derivative bulk action for the field $\phi$ these expressions become
\be
I=n_\mu\frac{\delta\call_{\mathrm{bulk}}}{\delta\partial_\mu \phi}\qquad\mbox{and}\qquad j=\frac{\delta \call_{\mathrm{DW}}}{\delta \phi}
\ee

Using the Lagrangian (\ref{matchaction}) in the case of our interest we can compute
\bea
I_T &=& n_\m {\d \call_{bulk} \over \d ( \pa_\m T )} = -2 (\m -1) \sqrt{-g}{ n^\m \pa_\m T \over T^2 - 4\p^2}\,,\\
I_C^i &=& n_\m {\d \call_{bulk} \over \d \pa_\m C_i}   = -2 (\m -1) \sqrt{-g} { n_\m G^{\m i} \over T^2 - 4\p^2}\,,\\
I_{A^L}^i &=& 2 n_\m {\d \call_{bulk} \over \d \pa_\m  {A^L}_i } = -{4 \over l} n_\m \e^{\m ij} {A^L}_j\,. 
\eea
The factor 2 in the last line arises because because the CS-Lagrangian is first order in derivatives and hence both the $A^L$ variation and the $\pa_\m  {A^L}$ variation of the bulk action give rise to a delta-function term with equal coefficient.
For the unit normal we concretely take \be
n_+ = {1\over \sqrt{ g_{UU} }} \pa_U\,,\qquad\qquad
n_- =  {1\over \sqrt{ g_{rr} }} \pa_r\,. \ee
The source terms from the domain wall are
\bea
j_T &=&  {\d \call_{DW} \over \d T } = 2  l (\m -1) \r \,,\\
j^i_C &=& {\d \call_{DW} \over \d C_i }=   2   (\m -1) \r \d^i_t\,,\\
j^i_{A^L} &=& {\d \call_{DW} \over \d {A^L}_i }= - 4 \p  \r b \d^i_t\,.
\eea
Computing the inside ($-$) and outside ($+$) values for the $I$'s, and the explicit form of the $j$'s from (\ref{inside}) and (\ref{outside}) we can see that the matching equations (\ref{matcheq}) are satisfied for each of the fields. So we see that this more detailed analysis reproduces the result obtained by the argument based on charge conservation.
\end{appendix}

\bibliographystyle{JHEP}
\bibliography{godel}

\end{document}